# Untangling individual cation roles in rock salt high-entropy oxides

*Saeed S. I. Almishal[1] and *Jacob T. Sivak[2], George N. Kotsonis[1], Yueze Tan[1], Matthew Furst[1], Dhiya Srikanth[1], Vincent H. Crespi[1,2,3], Venkatraman Gopalan[1], John T. Heron[4], Long-Qing Chen[1], *Christina M. Rost[5], Susan B. Sinnott[1,2,6], and Jon-Paul Maria[1]

[1]*Department of Materials Science and Engineering, The Pennsylvania State University, University Park, PA 16802, USA*

[2]*Department of Chemistry, The Pennsylvania State University, University Park, PA 16802, USA*

[3]*Department of Physics, The Pennsylvania State University, University Park, PA 16802, USA*

[4]*Department of Materials Science and Engineering, University of Michigan, Ann Arbor, MI, 48109, USA*

[5]*Department of Materials Science and Engineering, Virginia Polytechnic Institute and State University, Blacksburg, VA 24061, USA*

[6]*Institute for Computational and Data Sciences, The Pennsylvania State University, University Park, PA 16802, USA*

**Corresponding Authors:** Saeed S. I. Almishal saeedsialmishal@gmail.com, Jacob T. Sivak jts6114@psu.edu, and Christina M. Rost cmrost@vt.edu

## Abstract

We unravel the distinct roles each cation plays in phase evolution, stability, and properties within $Mg_{1/5}Co_{1/5}Ni_{1/5}Cu_{1/5}Zn_{1/5}O$ high-entropy oxide (HEO) by integrating experimental findings, thermodynamic analyses, and first-principles predictions. Our approach is through sequentially removing one cation at a time from the five-component high-entropy oxide to create five four-component derivatives. Bulk synthesis experiments indicate that Mg, Ni, and Co act as rock salt phase stabilizers whereas only Mg and Ni enthalpically enhance single-phase rock salt stability in thin film growth; synthesis conditions dictate whether Co is a rock salt phase stabilizer or destabilizer. By examining the competing phases and oxidation state preferences using pseudo-binary phase diagrams and first-principles calculations, we resolve the stability differences between bulk and thin film for all compositions. We systematically explore HEO macroscopic property sensitivity to cation selection employing both predicted and measured optical spectra. This study establishes a framework for understanding high-entropy oxide synthesizability and properties on a per-cation basis that is broadly applicable to tailoring functional property design in other high-entropy materials.

## Introduction

Modern materials development and design advancements present unorthodox methods where traditional thermodynamic routes like enthalpy no longer play the central role in phase stabilization. The seminal report on the prototypical high-entropy oxide (HEO) $Mg_{1/5}Co_{1/5}Ni_{1/5}Cu_{1/5}Zn_{1/5}O$ in 2015 exemplifies this paradigm in which crystalline oxide solutions are enabled by configurational entropy at sufficiently high temperatures [1]. At temperatures lower than approximately 875 °C, the solid solution state is unstable with respect to the formation of various compound oxide phases as illustrated in the left portion of Figure 1. When heated beyond 875 °C, near-equilibrium solid-state diffusion and reaction occur to entropy-stabilize a homogeneous compositionally disordered rock salt solid solution. Rapid cooling from high temperature kinetically traps a high-entropy solid solution configuration as a metastable state as illustrated in the right portion of Figure 1. Slow-cooling from high-temperature equilibrium or reannealing at intermediate or low-temperature results in phase separation as the system is kinetically allowed to approach equilibrium [1,2].

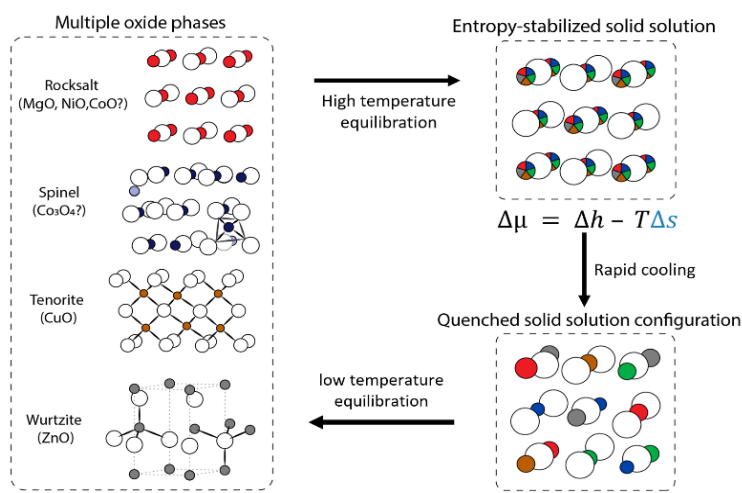

**Figure 1.** Flowchart illustrating phase progression in the $Mg_{1/5}Co_{1/5}Ni_{1/5}Cu_{1/5}Zn_{1/5}O$ system with temperature - reproduced from our JAcers feature article [2]. We highlight in this work that in addition to CuO and ZnO, $Co_3O_4$ should be explicitly considered in low temperature processes.

HEOs have garnered significant interest due to their remarkable chemical tunability, i.e., the ability to alter chemical composition while maintaining the same phase structure [2]. Rost et al. demonstrates that changing the concentration of a given end member in the $Mg_{1/5}Co_{1/5}Ni_{1/5}Cu_{1/5}Zn_{1/5}O$ system is possible, at the expense of the solid solution stability with an increased single-phase transition temperature. For instance, increasing the CuO fraction from 0.2 to 0.3 causes the single-phase stability temperature to rise from 875 °C to roughly 1050 °C [1]. Removing a single end-member should increase the single-phase stability temperature even more dramatically, if one exists at all prior to melting under equilibrium synthesis approaches [2]. Importantly, each cation is different in its "excess chemical energy" that it contributes to the homogeneous high-entropy solid solution. It is reasonable to expect that removal of certain cations such as Cu should be less detrimental to rock salt single-phase stability than others (Mg or Ni) in the $Mg_{1/5}Co_{1/5}Ni_{1/5}Cu_{1/5}Zn_{1/5}O$ system, however the high-entropy composition space results in nontrivial cation contributions that require a methodical exploration.

Alternative pathways beyond conventional equilibrium-based methods, such as solid-state synthesis, offer a rich and promising research avenue to stabilize more chemically diverse and/or non-equimolar HEO phases. The non-equilibrium kinetics enabled by physical vapor deposition (PVD) techniques, particularly pulsed laser deposition (PLD), rapidly condense precursors from a high-entropy initial state on a substrate trapping them in a metastable homogenous solid solution thin-film. In essence, kinetic stabilization allows access to a broader spectrum of *metastabilized* atomic and electronic configurations whose bulk synthesis may require extreme physical or chemical conditions[2–6]. Substrate temperatures are considerably lower than the bulk single-phase stabilization temperatures, thereby introducing an additional layer of complexity to the role that each cation plays in the stability of the system under thin-film growth boundary conditions. We find this particularly noteworthy in the case of the Co cation as it can adopt multiple oxidation states across our synthesis temperatures depicted in Figure 1 and Figure S1. This underscores the need for a comprehensive understanding of the various factors that govern the stability, and therefore also functionality, of these complex material systems.

HEOs disordered chemical makeup results in properties that oftentimes defy conventional predictions from their elemental constituents. These unique and potentially tunable properties are the result of synergistic interactions between compositional flexibility, disorder, entropic cation distribution, unusual cation coordination, and microstructural metastability. By carefully selecting elemental compositions and their relative concentrations, the community can manipulate atomic-scale properties to control macroscopic phenomena like electrocatalysis, magnetism, metal-insulator transitions, and even emergent topological responses [2,7–12]. Optical properties are particularly sensitive to local distortions, making them a useful tool for probing the underlying electronic structure of HEOs, particularly those that are insulating in character. HEOs have been shown to have small optical band gaps and broad refractive index features due to electronic disorder and broad total densities of states [2,4,13–16]. These characteristics make HEOs attractive for numerous applications such as photocatalysts [2,14]. Although there is exciting potential for continued property advancement, the synergetic relationship each cation plays on resulting properties still awaits exploration and implementation [2].

We therefore present a set of experiments designed to gain a deeper understanding of each cation's contribution to rock salt HEO stability and optical properties. We employ an element removal approach: each element is sequentially removed from the parent $Mg_{1/5}Co_{1/5}Ni_{1/5}Cu_{1/5}Zn_{1/5}O$, and the resulting response is observed, allowing us to unravel the distinct role each cation plays within $Mg_{1/5}Co_{1/5}Ni_{1/5}Cu_{1/5}Zn_{1/5}O$. We divide our analysis into three distinct sections: (1) stability of bulk ceramics sintered at relatively high temperatures (>875 °C), (2) stability of thin films grown with PLD at nonequilibrium synthesis conditions with relatively low substrate temperatures (<600 °C), and (3) functional response of linear optical properties. In the first two sections, we complement experimental findings with thermodynamic analysis enabled by first-principles enthalpy calculations and CALPHAD pseudo-binary phase diagrams. In the third section, however, calculated electronic structures are analyzed prior to experimental results. This investigation contributes a more comprehensive understanding of the phase evolution, stability, and properties of the prototypical entropy-stabilized rock salt, as well as establishes an experimental and computational framework to investigate the role each cation contributes to both synthesizability and functionality of other entropy-driven materials.

# Main

*1. Cation selection trends in single-phase formation under equilibrium synthesis*

It is advantageous to first revisit the phase evolution for the parent $Mg_{1/5}Co_{1/5}Ni_{1/5}Cu_{1/5}Zn_{1/5}O$ five-component phase to realize the role that each cation plays in rock salt HEO stability. In Figure 2, we reproduce the original plot by Rost et al.[1] with a longer sintering time profile. The temperature spanned a range from 700 to 900 °C in 50 °C increments. After 750 °C, four prominent phases are observed: rock salt, tenorite, spinel, and wurtzite. The more pronounced secondary phases that we observe compared to the original work are the result of increased sintering (i.e, equilibration) time. Importantly, reversibility still holds, a key requirement of entropy-driven transitions. The transformation from single phase at 900 °C to multiphase at 750 °C, to single phase at 900 °C is evident by the X-ray patterns and demonstrates the enantiotropic phase transition.

To better rationalize the phases present and their evolution in the parent HEO diffraction patterns in Figure 2, we utilize the rich thermodynamics offered by the constituent phase diagrams. Conventional binary phase diagrams provide unique insights into the complex high-entropy system phase evolution with temperature and provide phase maps to navigate enthalpy-driven competing phases. We generate, using CALPHAD, the 10 binary phase diagrams derived from $Mg_{1/5}Co_{1/5}Ni_{1/5}Cu_{1/5}Zn_{1/5}O$ in Figures S2-S6 with the x-axis being the increase in concentration for the cation less prone to stabilize in the rock salt structure (details in methods). These phase diagrams convey the differing complexity of thermodynamic landscapes for each cation. The simplest are MgO and NiO which are completely soluble in the rock salt structure and thus can be considered *rock salt phase stabilizers*. The role Cu, Zn, and Co play in high-entropy rock salt phase stability is more convoluted, however, especially if entropic contributions are considered.

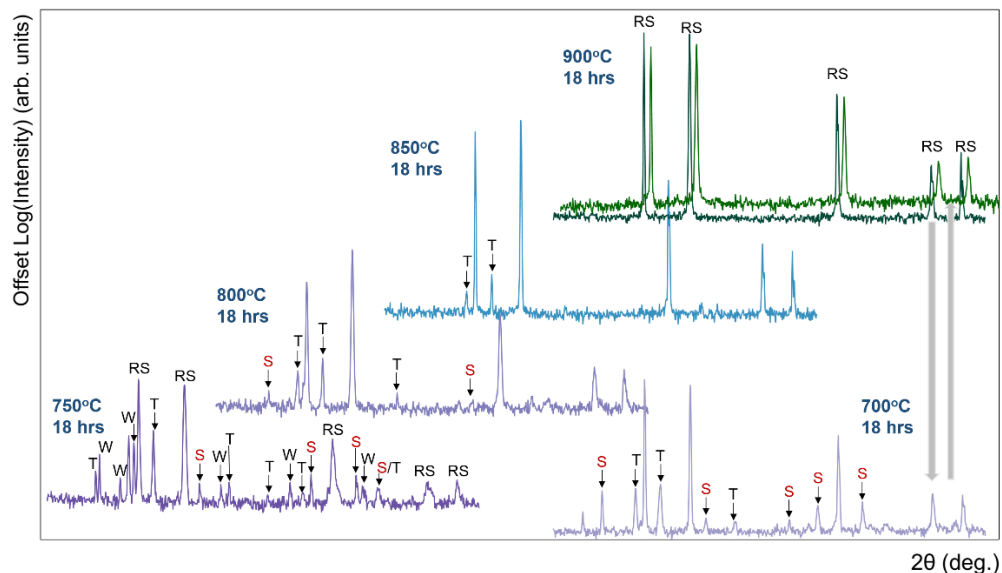

**Figure 2.** X-ray diffraction patterns on a logarithmic scale for equimolar mixture of MgO, NiO, ZnO, CuO and CoO to form $Mg_{1/5}Co_{1/5}Ni_{1/5}Cu_{1/5}Zn_{1/5}O$. The patterns were collected from four pellets. The three first pellets were equilibrated for 18 hours at different temperatures (750 °C, 800 °C and 850 °C) in air, then air quenched to room temperature by direct extraction from the furnace. The fourth pellet was equilibrated for 18hrs and consequently quenched three times as indicated by the arrows (900 °C - 700 °C - 900 °C). Arrows and letters indicate peaks associated with non-rock salt phases, peaks indexed with (T), (W), (SP) and with (RS) correspond to tenorite, wurtzite, spinel, and rock salt phases, respectively. The two x-ray patterns for 900 °C annealed samples are offset in 2θ for clarity.

CuO strongly favors the tenorite structure and exhibits limited solubility in the rock salt phase (Figures S5-S6). Inspecting the phase diagrams also reveals that CuO has the propensity to form ionic compounds with MgO and NiO, has the earliest melting point compared to the four other cations, and reduces to cuprite $Cu_2O$ before melting. Supported by our current findings and numerous other synthesis reports, tenorite CuO is the last phase to be incorporated into the rock salt matrix as well as the initial secondary phase to precipitate upon temperature annealing [1,5,17–19]. Consequently, we regard CuO as a *rock salt phase destabilizer*, which requires mixing with rock salt stabilizers such as Mg and Ni to form a single-phase rock salt. We note that cations identified as rock salt destabilizers are particularly interesting from an entropy-stabilization perspective as they may provide unique local structures with untapped property opportunities[5].

ZnO favors the wurtzite structure at our synthesis conditions. Yet, inspection of the MgO-ZnO and NiO-ZnO phase diagrams reveals a nonnegligible solubility of Zn in MgO and NiO at elevated temperatures (Figure S4) [20]. Even at temperatures as low as 600 °C, around 20% Zn can be readily dissolved in the MgO or NiO rock salt matrix. This could explain the early disappearance of wurtzite phase peaks at 800 ºC in Figure 2, in contrast to the strong persistence of the tenorite phase until ~900 ºC. Wurtzite ZnO also does not appear to precipitate out of the rock salt solution upon 700 ºC annealing, while prominent CuO tenorite peaks emerge. Notably, we report here that equimolar $Mg_{1/3}Ni_{1/3}Zn_{1/3}O$ can be stabilized in the rock salt phase at ~1300 ºC in air, requiring a much higher temperature than the parent five-component HEO (~875 ºC) (Figure S7). ZnO thus requires an "extra" ~400 °C to dissolve into the ternary rock salt oxide matrix at 33% compared to the 20% dissolved in the quinternary rock salt oxide matrix. This result highlights another key entropy-stabilized oxide feature: *configurational entropy boosts the TS term in Gibbs free energy resulting in lower temperature barrier for phase transitions compared to less-component derivatives.*

Co readily adopts multiple oxidation states over relevant temperatures to form either rock salt CoO ($Co^{2+}$) or spinel $Co_3O_4$ (both Co2+ and $Co^{3+}$) in contrast to the four other cations, see Figure S1. The Co-O temperature-pressure phase diagram indicates that $Co_3O_4$ is the equilibrium phase below ~800 ºC in ambient pressure, while CoO is the stable phase at higher temperatures. MgO-CoO and NiO-CoO phase diagrams additionally reveal that above ~770 °C 20% Co dissolves into the rock salt solid solution, while lower temperatures result in a spinel-rock salt multi-phase mixture. Synthesis conditions therefore dictate whether Co acts as a *rock salt phase stabilizer* (relatively high temperatures) or *destabilizer* (low temperatures). Rock salt CoO was used as the starting powder for all pellets shown in Figure 2, yet a spinel phase (containing both $Co^{2+}$ and $Co^{3+}$) forms in $Mg_{1/5}Co_{1/5}Ni_{1/5}Cu_{1/5}Zn_{1/5}O$ targets sintered below 850 ºC. Spinel phase peaks are also evident following 700 ºC annealing of the quenched high-entropy single-phase rock salt. In all cases, no spinel peaks are observed at 850 ºC or above regardless of if the starting powder in the raw powder mix is CoO or $Co_3O_4$ (Figure S8). This affirms that CoO is the stable Co-containing phase at relatively high temperatures in the five-component high-entropy mixture, while $Co_3O_4$ will prevail at relatively lower temperatures. Markedly, heating quenched single-phase rock salt $Mg_{1/5}Co_{1/5}Ni_{1/5}Cu_{1/5}O$ in an inert medium (such as Ar or $N_2$) produces no spinel secondary phases, highlighting the $O_2$ chemical potential role in phase evolution [18]. Therefore, Co favorable oxidation state of at different synthesis conditions must be accounted for when assessing $Mg_{1/5}Co_{1/5}Ni_{1/5}Cu_{1/5}Zn_{1/5}O$ rock salt single-phase stability.

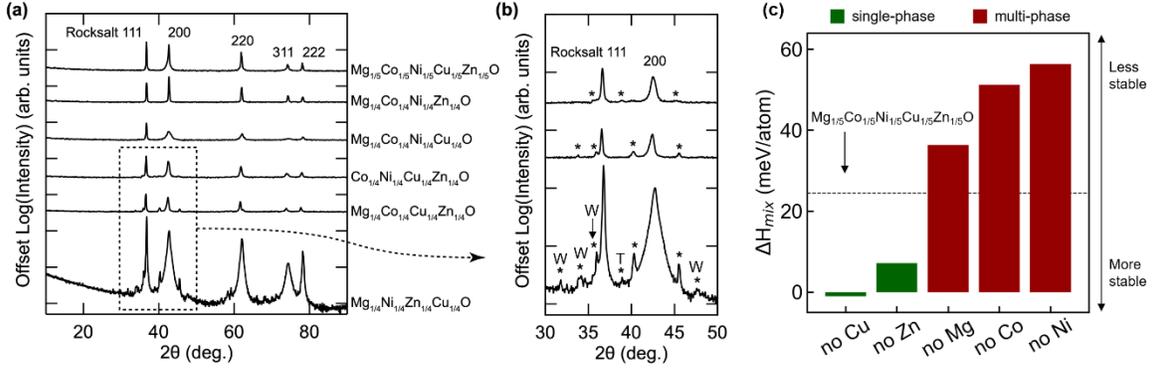

**Figure 3.** Bulk ceramic stability analysis. (a) X-ray diffraction patterns of $Mg_{1/5}Co_{1/5}Ni_{1/5}Cu_{1/5}Zn_{1/5}O$ parent and four-component derivatives. (b) Narrower angular range for derivative compositions containing secondary phase peaks – the unidentified secondary phase in the case without Co could be $Cu_3MgO_4$, while the secondary phase in the case without Ni could be $CoCu_2O_3$. (c) DFT stability analysis of four-component derivatives with respect to ground-state AO phases using mixing enthalpy. $Mg_{1/5}Co_{1/5}Ni_{1/5}Cu_{1/5}Zn_{1/5}O$ parent is shown as a dashed line.

The high-entropy rock salt phase stability can be further evaluated by sequentially removing each cation from the parent $Mg_{1/5}Co_{1/5}Ni_{1/5}Cu_{1/5}Zn_{1/5}O$ to form five unique four-component derivatives. In their original work, Rost et al. showed that none of the four-component derivative set forms a single-phase rock salt at 875 °C (the threshold temperature for parent HEO complete solubility), supporting that the five-component system is entropy-stabilized. Here in Figure 3, we show bulk ceramic XRD patterns of these derivatives sintered at 1000 °C to form a comparison between cations at the highest temperature before CuO reduction and melting onsets. The predominant phase of each derivative is a rock salt solid solution. Removing Cu or Zn results in single-phase rock salt solid solution at 1000 °C. However, removing Mg, Co, or Ni greatly increases the temperature needed to stabilize the rock salt phase beyond 1000 °C as evidenced by the secondary phases present in the XRD (Figure 3b). Importantly, these results support our thermodynamic hypothesis that Co acts as a rock salt phase stabilizer in addition to Mg and Ni at relatively higher temperatures (>800 °C). HEOs inherent local distortions generally result in broad XRD peaks. We attribute the particularly broader peaks for the case without Zn ($Mg_{1/4}Co_{1/4}Ni_{1/4}Cu_{1/4}O$) to an increased Cu content in the rock salt matrix as Cu exhibits a Jahn-Teller distortion of its octahedra [21–23]. The greater signal-to-noise ratio in the bottom trace of Figures 3a and 3b is due to the removal of Co, which fluoresces Cu-Kα radiation and increases the noise floor when present in samples.

All constituent cations are expected to be 2+ for our bulk synthesis conditions as established by the thermodynamic analysis, hence only the ground-state $A^{2+}O^{2-}$ polymorphs are considered as competing phases to the high-entropy rock salt single-phase. DFT calculations are used to calculate the mixing enthalpy:

$$\Delta H_{mix} = E[HEO] - \sum_i n_i E[A_i^{2+}O^{2-}] \qquad (1)$$

where $i$ indicates individual constituents of the HEO, and $n_i$ is the fractional amount of each cation in the HEO – we use HEO nomenclature following our definition in ref. [2]. $\Delta H_{mix}$ can be considered as the enthalpic penalty to the single-phase rock salt HEO formation for bulk ceramic synthesis at 1000 °C in ambient conditions. $\Delta H_{mix}$ for all four-component derivatives shown in Figure 3c captures the same trends as the experiment: Mg, Co, or Ni are rock salt stabilizers as their removal increases $\Delta H_{mix}$, while removal of Cu or Zn results in a decreased $\Delta H_{mix}$, making these cations rock salt destabilizers at these boundary conditions. Calculated $\Delta H_{mix}$ indicates that Cu destabilizes the rock salt phase to a greater extent than Zn, matching synthesis results that CuO tenorite is more 'stubborn' resisting dissolution into the rock salt matrix until higher temperatures are accessed. $\Delta H_{mix}$ values for the compositions where a rock salt stabilizer has been removed (no Mg, no Co, and no Ni) indicate that Co and Ni stabilize the HEO rock salt phase more than Mg. These findings are consistent with the bulk ceramic X-ray diffraction in which Co or Ni removal results in more apparent, persistent secondary phases compared to Mg removal (Figure 3b), highlighting the effectiveness of first-principles calculations for exploring HEO stability.

*2. Cation selection trends in single-phase formation under nonequilibrium synthesis*

Exploring non-equilibrium routes to *metastabilize* $Mg_{1/5}Co_{1/5}Ni_{1/5}Cu_{1/5}Zn_{1/5}O$ and derivative phases is an intriguing proposition to probe relatively low-temperature stability and cation solubility limit. X-ray diffraction patterns for the four-component derivative thin films grown at substrate temperatures from 200 °C to 600 °C are shown in Figures 4a-e. Compositions are ordered from left to right based on the substrate-temperature range over which epitaxial single-phase rock salt films can be grown. Figures 4f-j illustrate higher resolution X-ray diffraction patterns for the subset of films identified as single-phase with relatively good crystalline quality. All the films in this subset are commensurate with the MgO substrate as confirmed by the reciprocal space maps (RSMs) included in the supporting information (Figures S9-S12). In all cases except for removing Cu, the single-phase growth temperature window shrinks relative to the five-component parent compound. Notably, removing Ni results in the smallest growth window with a single phase only realized at 300 °C growth, corroborating our thermodynamic and DFT observations that Ni is a rock salt stabilizer while Cu is a destabilizer with the highest tendency to depart from the high-entropy phase. In a recent study, we have also reported that if films are deposited with a 'slow' growth rate, CuO dissolutes from the high-entropy matrix forming coherent CuO nanotweeds that imitates CuO precipitation in bulk synthesis on a much smaller length scale (not detected in x-ray diffraction) [5].

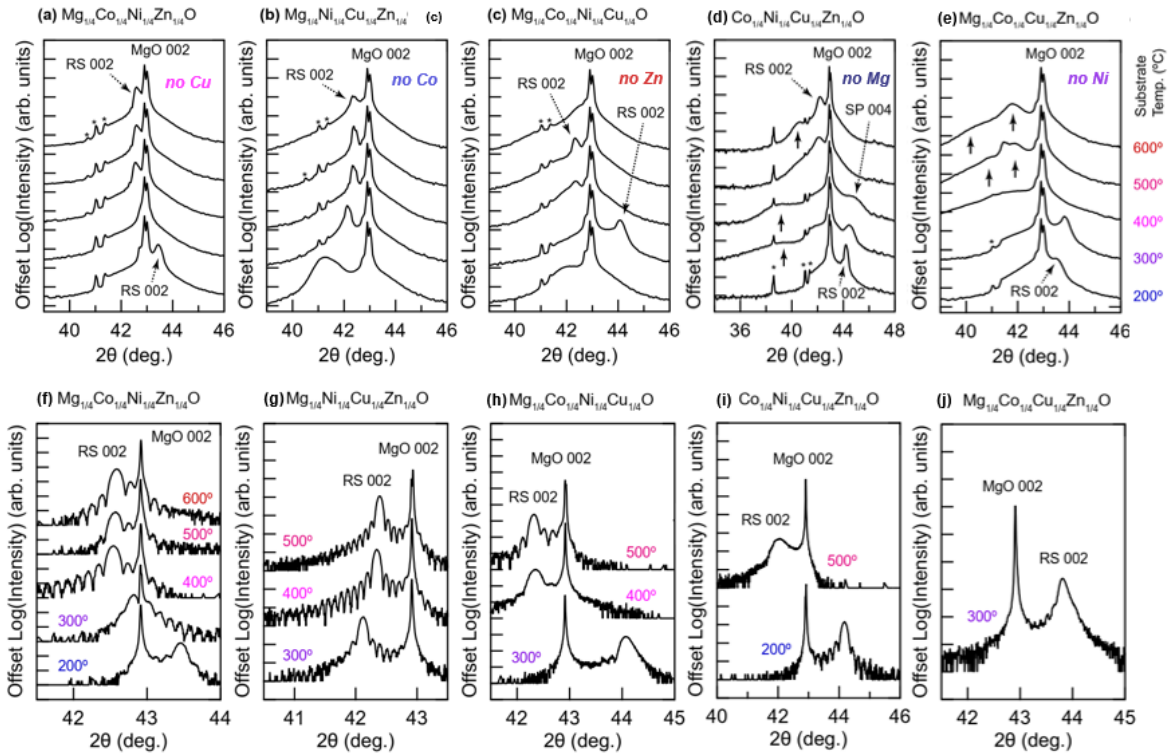

**Figure 4.** X-ray diffraction patterns of $Mg_{1/5}Co_{1/5}Ni_{1/5}Cu_{1/5}Zn_{1/5}O$ and four-component derivative thin films. (a-e) are low resolution scans with divergent beam for all films grown at temperatures 200 °C to 600 °C and (f-j) are higher resolution scans with parallel beam for only single phase films. Asterisks indicate Cuβ and W reflections.

The change in lattice parameter and epitaxial strain state across the investigated substrate temperatures for the cases we remove Cu, Zn or r Co ($Mg_{1/4}Co_{1/4}Ni_{1/4}Zn_{1/4}O$, $Mg_{1/4}Co_{1/4}Ni_{1/4}Cu_{1/4}O$, or $Co_{1/4}Ni_{1/4}Cu_{1/4} Zn_{1/4}O$) aligns with the trend our team previously reported for the parent phase $Mg_{1/5}Co_{1/5}Ni_{1/5}Cu_{1/5}Zn_{1/5}O$ thin films [4]. At low growth temperatures, the unit cell volume is smaller than the MgO substrate resulting in tensile epitaxial strain. We attribute this trend to the Co tendency for 3+ at low growth temperatures and 50 mTorr $O_2$ (as can be seen from Co-O phase diagram in Figure S1). Conversely, growth at higher substrate temperatures results in an enhanced Co2+ concentration and unit cell volume increase beyond that of MgO, causing compressive epitaxial strain in the films. Removing Co precludes the structural trend as shown in Figure 4b and 4g, confirming that Co is the primary contributor. The unit cell volume increase is also the least abrupt for the case with no Cu ($Mg_{1/4}Co_{1/4}Ni_{1/4}Zn_{1/4}O$) suggesting that the local anisotropy associated with Jahn-Teller distorted $CuO_6$ octahedra [21,22] may contribute to the structural transition abruptness as a function of substrate temperature. Based on these collective observations, Co exhibits a propensity similar to Cu to exsolve from the metastabilized rock salt matrix. These findings are in line with our recent study, where we have reported the strain-mediated formation of $Co^{3+}$-rich spinel nanocuboids in $Mg_{1/5}Co_{1/5}Ni_{1/5}Cu_{1/5}Zn_{1/5}O$ thick films grown at 400 °C with large strain energy per unit volume [5].

A particular case to consider is $Co_{1/4}Ni_{1/4}Cu_{1/4} Zn_{1/4}O$ (no Mg) which appears to grow with a single-phase rock salt structure at 200 °C and 500 °C with opposite strain. At intermediate temperatures, it grows with a two-phase structure of rock salt and what could be spinel, see Figure 4d. For a more thorough analysis, we show in Figure 5a a complementary set of $Co_{1/4}Ni_{1/4}Cu_{1/4}Zn_{1/4}O$ (no Mg) films grown with a [111] orientation on [00.1]-$Al_2O_3$. The films exhibit a similar lattice parameter trend as the films grown on MgO. The film grown at 400 °C confirms the formation of spinel as a secondary phase which can again be attributed to the propensity for $Co^{3+}$ to form at our synthesis conditions. The rock salt diffraction pattern for $Co_{1/4}Ni_{1/4}Cu_{1/4}Zn_{1/4}O$ (no Mg) grown at 200 °C on MgO and [00.1]-$Al_2O_3$ may indicate that a structure with rock salt symmetry nucleates initially with significant Co3+ concentration and can be quenched with a low enough growth temperature. We hypothesize the quenched rock salt may then include significant amounts of tetrahedral cations and vacant octahedral lattice sites, but the kinetics evidently do not allow the system to chemically configure into a crystal with observable spinel symmetry via XRD. In the case of 400 °C substrate temperature, however, Co3+ has more energy to evolve during growth into observable spinel symmetry. This is consistent with our observations of spinel nanocuboids in the parent phase grown thick at 400 °C [5]. The rock salt structure observed after 500 °C growth, with a larger lattice parameter, is consistent with primarily 2+ cations in the structure.

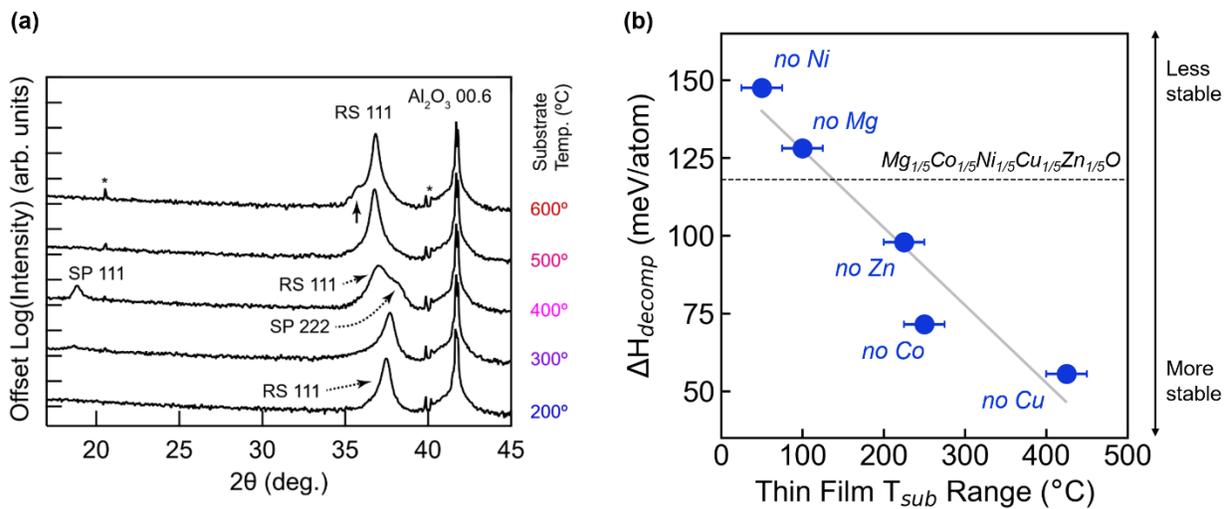

**Figure 5.** (a) X-ray diffraction patterns of $Co_{1/4}Ni_{1/4}Cu_{1/4} Zn_{1/4}O$ (no Mg) films on [00.1]-Al2O3 as a function of temperature, (b) Single-phase substrate temperature range for rock salt thin films plotted against the DFT calculated decomposition enthalpy. Linear fit is shown as a grey line, while the five-component parent decomposition enthalpy is plotted as a black dashed line.

While we have shown that Co is a rock salt phase stabilizer for bulk ceramics, Co interestingly acts against homogeneous rock salt phase formation in thin films. This is evident by the wide temperature range where a single-phase rock salt can be grown upon Co removal compared to removal of other cations, as well as by the spinel phase formation under various synthesis conditions. Following our thermodynamic argument, this is likely a result of its propensity for 3+ valence at lower temperatures (which includes all substrate temperatures explored). DFT provides an accompanying understanding to our thermodynamic argument through the decomposition enthalpy ($\Delta H_{decomp}$), which considers *all* ground-state polymorphs across the relevant chemical landscape as competing phases:

$$\Delta H_{decomp} = E_{rxn} = E_{ABC} - E_{A-B-C}. \qquad (2)$$

$\Delta H_{decomp}$ consequently considers the possibility of multiple valence states (assuming these lead to a lower competing reaction energy). This is in contrast to $\Delta H_{mix}$ we used for bulk ceramic stability which only considers 2+ cations. $\Delta H_{decomp}$ is thus considered as a measure of instability with respect to phase separation [24,25], which can be used to evaluate our HEO thin films grown at relatively low temperatures. $\Delta H_{decomp}$ for $Mg_{1/5}Co_{1/5}Ni_{1/5}Cu_{1/5}Zn_{1/5}O$ and its four-component derivatives are shown in Figure 5b, where removal of Ni again results in the most unstable formulation and, conversely, removal of Cu is the most stable. Co indeed acts as a rock salt destabilizer in $\Delta H_{decomp}$ calculated with DFT, as its removal lowers the enthalpic cost compared to the parent $Mg_{1/5}Co_{1/5}Ni_{1/5}Cu_{1/5}Zn_{1/5}O$. For all compositions containing Co, a spinel phase is present in the decomposition reaction (Table S2), where the 3:4 cation-to-oxygen ratio necessitates 2/3 of the cations having a 3+ valence. This supports our thermodynamic argument that spinel $Co_3O_4$ is favored over a rock salt CoO during HEO thin film growth due to the lower substrate temperatures compared to the relatively high temperatures of bulk ceramic synthesis.

$\Delta H_{decomp}$ values are compared to the substrate temperature range that four-component derivatives can be grown as a single-phase rock salt. A linear correlation is found between these values, further confirming that the decomposition enthalpy is an appropriate quantity for evaluating the enthalpic cost for single-phase rock salt HEO formation at lower temperature conditions. The order in which secondary phases dissolve into the parent rock salt HEO with increasing temperature in Figure 2

(ZnO wurtzite – $Co_3O_4$ spinel – CuO tenorite) remarkably aligns with decreasing $\Delta H_{decomp}$ quantities for the rock salt destabilizers (Zn – Co – Cu). This suggests that decomposition enthalpies may also qualitatively capture the order in which secondary phases fully dissolve in the rock salt matrix with temperature.

*3. Cation selection trends in functionality and linear optical properties*

Our collective analysis indicates that synthesis conditions dictate whether a cation will assist or hinder the rock salt HEO structure, both of which may provide notable property opportunities. We focus solely in this section on samples which could be grown as high-quality epitaxial single-phase thin films. Insights into the role each cation plays in functional tunability of $Mg_{1/5}Co_{1/5}Ni_{1/5}Cu_{1/5}Zn_{1/5}O$ and its derivatives are first provided by first-principles electronic structure calculations. The predicted density of states for all HEOs are shown in Figure 6a where our calculated electronic band gap of the parent $Mg_{1/5}Co_{1/5}Ni_{1/5}Cu_{1/5}Zn_{1/5}O$ is 0.48 eV, in good agreement with the experimentally reported value of ~0.8 eV [26]. Better band gap agreement could be achieved through the addition of a Hubbard U correction [27], however we rely on the predictions from the metaGGA $r^2$SCAN functional alone. The valence band maxima is dominated by Cu and Co d-orbitals, while Cu d-orbitals predominantly account for the conduction band minima. For clarity, only these orbitals and the total density of states are shown. Cu therefore controls the HEO electronic structure landscape near the Fermi energy such that its removal results in a dramatic increase in electronic band gap (1.33 eV) compared to all other compositions in this study (~0.4 eV). These observations are also reflected in the calculated optical absorption spectra (Figure S13), where $Mg_{1/4}Co_{1/4}Ni_{1/4}Cu_{1/4}O$ (no Cu) absorption in the visible regime is greatly reduced compared to its four-component counterparts and five-component parent.

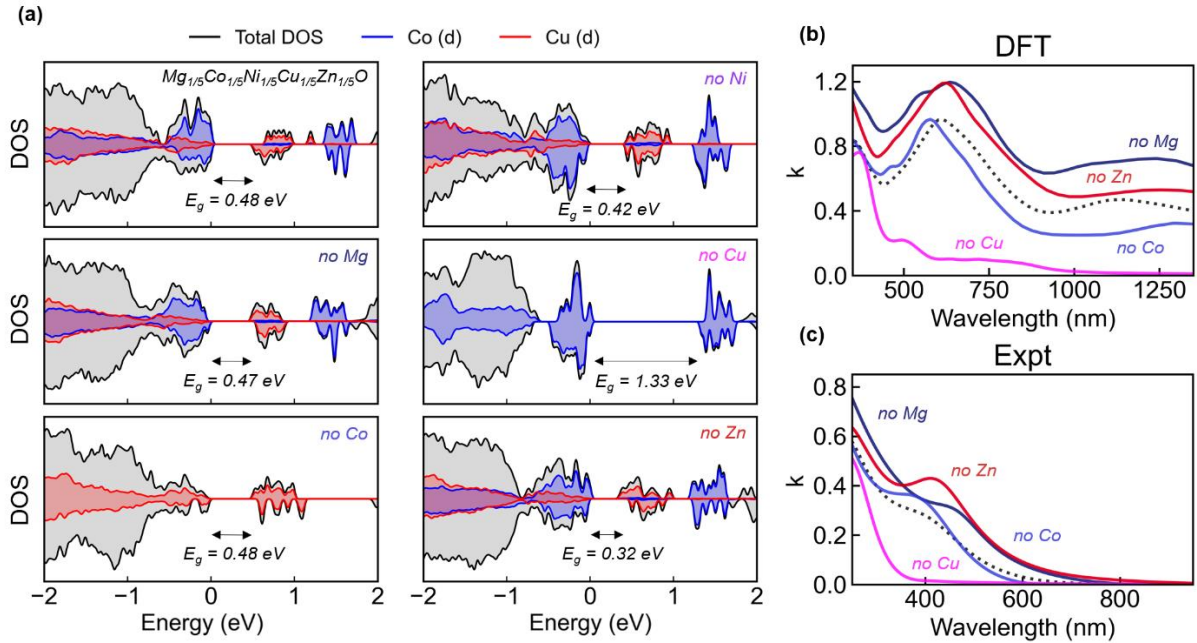

**Figure 6.** (a) DFT calculated density of states for $Mg_{1/5}Co_{1/5}Ni_{1/5}Cu_{1/5}Zn_{1/5}O$ and four-component derivatives. The Fermi energy is shifted to 0 eV in all cases. **(b)** DFT calculated imaginary portion of the refractive index (k) using the independent particle approximation and **(c)** experimentally measured imaginary portion of the refractive index. Note the difference in wavelengths between DFT predictions to better align with experimentally measured spectra. $Mg_{1/4}Co_{1/4}Cu_{1/4}Zn_{1/4}O$ (no Ni) is not included as it could only be grown at 300°C as a single-phase film.

Early dielectric characterization of $Mg_{1/5}Co_{1/5}Ni_{1/5}Cu_{1/5}Zn_{1/5}O$ and derived rock salts show large and tunable permittivity dispersions [26]. Large loss tangents at low frequencies and high temperatures indicate space charge contributions, potentially from mixed ionic–electronic conductivity in related materials. These early observations have prompted further development of HEO dielectrics [28,29]. We pinpoint the complex refractive index real (n) and imaginary (k) parts as particularly sensitive indicators for discerning optical property responses as a function of cation selection. The imaginary part (k) is utilized as an optical absorption indicator given the relation:

$$\alpha = \frac{4\pi k}{\lambda}. \qquad (3)$$

Cation selection indeed plays a significant role in the optical properties of HEOs as observed in Figure 6b for DFT calculated complex refractive indices of the parent $Mg_{1/5}Co_{1/5}Ni_{1/5}Cu_{1/5}Zn_{1/5}O$ and its derivatives. In Figure 6c we show the corresponding experimental measurements for films grown at 500 °C as these are primarily stoichiometric single-phase rock salt with limited $Co^{3+}$ formation. A different wavelength region is plotted for the DFT predictions to better align with experimental spectra due to the expected DFT band gap underestimation [30,31]. All compositions except that without Cu ($Mg_{1/4}Co_{1/4}Ni_{1/4}Zn_{1/4}O$) exhibit a feature in k around 650 nm in Figure 6b, indicating a resonant absorption. The analogous feature is measured experimentally around 400 nm. Our calculated density of states allows us to attribute this feature to the presence of Cu d-orbitals which occupy the bottom of the conduction band (Figure 6a). Additionally, it is important to note that qualitative features of the experimental spectra in Figure 6c are captured with DFT using only the simplified and relatively fast independent particle approximation (IPA). This showcases that we can potentially screen for desirable optical properties in different HEO formulations using computationally accessible techniques.

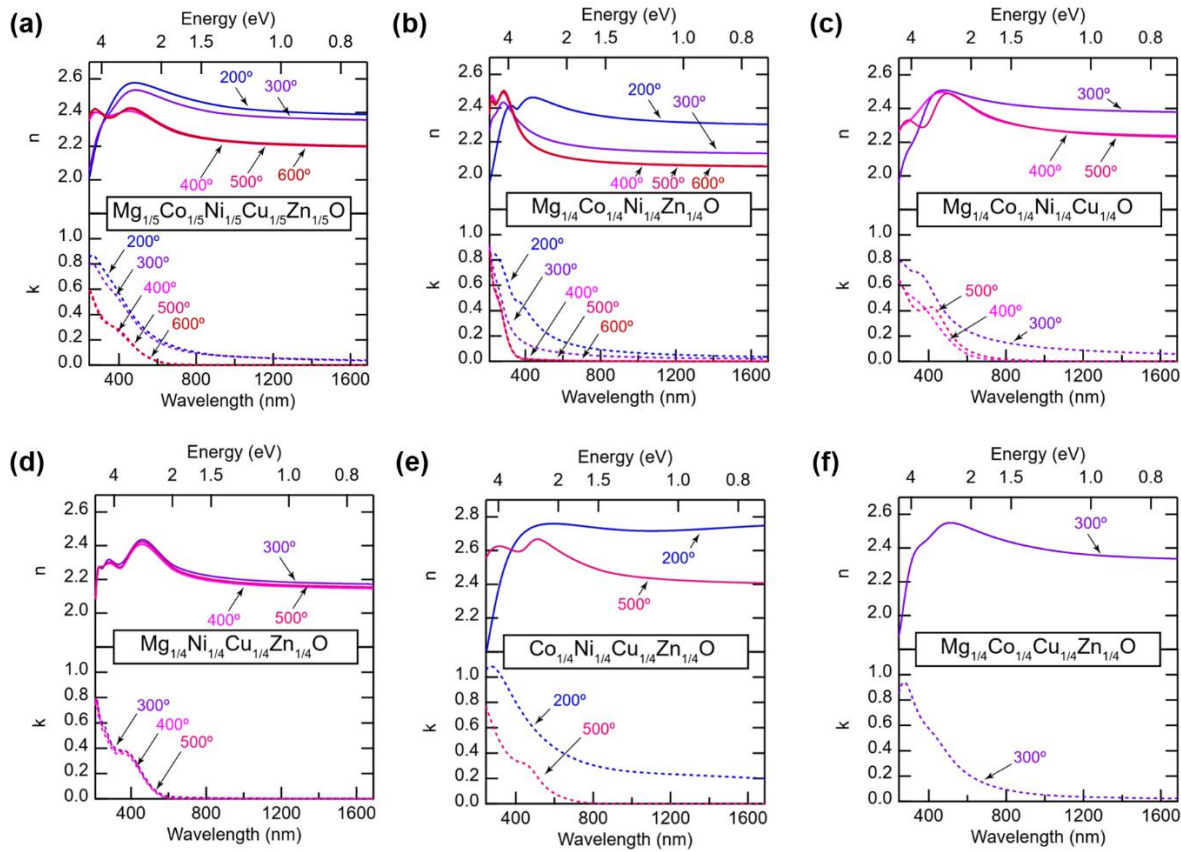

**Figure 7.** Complex refractive index models for $Mg_{1/5}Co_{1/5}Ni_{1/5}Cu_{1/5}Zn_{1/5}O$ and four-component derivative thin films, all with the same thickness. Substrate temperature during thin film growth is indicated near each trace. Note the trend of increasing *n* and *k* with growth at lower temperatures for each composition except $Mg_{1/4}Ni_{1/4}Cu_{1/4}Zn_{1/4}O$ (no Co).

Experimentally measured complex refractive index models for Figure 4i-j single-phase and epitaxial thin films as shown in Figure 7 to elucidate the effect that growth temperature has on optical functionality. Both the real ($n$) and the imaginary part ($k$) components of the refractive index are shown for a more encompassing look into the optical sensitivity of each HEO composition. The $k$ plots show a consistent pattern: lower growth temperatures lead to a stronger absorption tail leading to a smaller optical bandgap. This matches what is observed visually in Figure S14: films grown at lower temperatures appear darker and less transparent compared to the films grown at relatively higher temperatures. We link this increase in absorption to an enhanced $Co^{3+}$ concentration at lower temperatures growth. It follows that there is no significant change in optical properties with temperature in films without Co (Figure 7d), further confirming Co is primarily responsible for the temperature trend. It is important to acknowledge the limitations of the present optical models. In particular, all models assumed optical isotropy and fully dense microstructures with no optical property gradients. These assumptions were made to minimize the number of fitting parameters and provide self-consistent data sets with acceptable mean square error (MSE) values. Since our model fitting parameters were the same and consistent across all samples, we feel confident qualitatively comparing the models to identify trends as a function of synthesis conditions and cation selection. However, a point of future work is to develop more rigorous models that account for microstructure and the potential influence of epitaxial strain.

Predicted and measured optical data collectively indicate that adding or removing cations alters the range of accessible ion configurations, which in turn controls the available spectrum of electronic states that determine the optical response. Cation selection coupled to PLD growth temperature thus provides a viable route for tuning optical properties and potentially other functionalities within rock salt HEO materials.

## Summary and Outlook

We demonstrate how cation selection can be tailored to achieve phase stability and design physical and functional properties in a predictive manner. Integrating experimental observations with thermodynamic analysis and computational modeling enhance our understanding of these complex systems [32,33]. We emphasize that pseudo-binary phase diagrams provide abundant information about complex HEOs cation solubility, competing phases, and synthesizability. Stability trends on a per-cation basis for both bulk and thin film synthesis are resolved with DFT enthalpy calculations when competing phases are properly considered. This integrated investigative approach to quantify high-entropy synthesizability using computation and experiment enables accelerated entropy-driven materials discovery.

The tendency for a multi-phase solid solution upon Mg and Ni removal in both bulk and thin films suggests these cations are rock salt phase stabilizers in $Mg_{1/5}Co_{1/5}Ni_{1/5}Cu_{1/5}Zn_{1/5}O$, with Ni being a stronger stabilizer than Mg. Removing Cu and Zn has the opposite effect making these cations rock salt phase destabilizers, with Cu destabilizing the rock salt phase more than Zn. Co has a more complex behavior: although it drives single-phase rock salt HEO formation for bulk at T > 875°C, it exsolutes from the homogenous rock salt thin films at T < 875°C when given enough time in a relatively oxygen-rich environment. This complexity in Co behavior provides promising structural opportunities for property design, namely, the lattice parameter trend with temperature [4] and nano-spinel phase formation [5].

While configurational entropy undoubtedly plays an integral role within HEOs to overcome enthalpic costs, other entropic contributions such as vibrational, electronic, and magnetic entropy, which are shaped by cation selection, govern resulting properties. This implies the importance of carefully selecting and examining the high entropy system chemical makeup. Further, the ability to metastabilize all $Mg_{1/5}Co_{1/5}Ni_{1/5}Cu_{1/5}Zn_{1/5}O$ four-component derivatives through PVD provides another focal implication: HEO solid solutions covering an expansive composition and property space between the four-component derivatives and the five-component parent phase can also be metastabilized. Consequently, the chemical tunability inherent to these materials, manifested in cation selection and proportions, should enable a multitude of new materials with vast possibilities for property discovery and implementation.

## Methods

*Bulk synthesis*

Bulk ceramics were mixed to give stoichiometric solid solutions from binary oxide powders: MgO (Sigma-Aldrich, 342793), CoO (Sigma-Aldrich, 343153), $Co_3O_4$ (Millipore-Sigma, 637025), NiO (Sigma-Aldrich, 203882), CuO (Alfa Aesar, 44663) and ZnO (BeanTown Chemical, 211845). The mix was shaker-milled with 5 mm diameter yttrium-stabilized zirconia milling media for 2.5 hrs. The powder mix was pressed uniaxially into a 2.5 cm diameter pellet at 140 MPa (Carver Laboratory Press). The pellets were sintered in air at 1000 °C for 12-18 hrs and air-quenched by direct extraction from the hot zone of the furnace.

*Thin-film growth*

For each composition, a series of films was grown using pulsed laser deposition (PLD) on MgO substrates at temperatures ranging from 200 to 600 °C. A laser pulse rate of 6 Hz was used, and the ablation atmosphere was 50 mTorr of flowing oxygen. The $Co_{1/4}Ni_{1/4}Cu_{1/4}Zn_{1/4}O$ (no Mg) films were simultaneously grown film on [00.1]-$Al_2O_3$ substrate. X-ray diffraction (XRD) θ-2θ scans were collected using a PANalytical Empyrean diffractometer with Bragg-Brentano HD and PIXcel3D detector. We employed a 2-bounce Ge hybrid monochromator for the high-resolution scans and reciprocal space maps (RSMs).

*Optical analysis*

Spectroscopic ellipsometry was performed on a J.A. Woollam RC-2 ellipsometer. Scans were taken at incident angles of 50°, 60°, and 70°. Data was analyzed with the J.A. Woollam CompleteEASE software package assuming the films have uniform and isotropic optical constants. Data fitting employed two types of isotropic refractive index models. The first is an oscillator model that consisted of two Tauc-Lorentz oscillators and covered the entire wavelength range (about 200 to 1650 nm). In these models, thickness and roughness parameters were constrained to within a few nm of film thickness and roughness indicated by X-ray reflectivity measurements. The second set of models employed a simpler three-term isotropic Cauchy model for the real part of the refractive index as a function of wavelength for wavelength above 600 nm to avoid absorption features. The imaginary part of the refractive index was modeled with a simple exponential as a function of photon energy E (in eV) to account for an Urbach absorption tail[34]. All ellipsometry models have suitable figures of merit (MSE values of about 5 or less) and were developed using the minimum amount of fitting parameters necessary.

*Phase diagrams construction*

We employed OpenCalphad [35] and pycalphad [36] for the thermodynamic evaluations with prescribed thermodynamic data as described from Table S1. Chemical potential of solution phases are evaluated using the compound energy formalism [37], presuming pseudo-binary interactions for all solution phases. All pseudo-binary phase diagrams are calculated at $p(O_2) = 0.21$ bar. $CoO_x$ stability is evaluated with 0.001~200 Torr partial pressure of oxygen.

*Density functional theory calculations*

The Vienna Ab-initio Software Package (VASP) 6.4.1 is used for DFT calculations with the projector augmented wave pseudopotentials v54. The regularized-restored strongly constrained and appropriately normed ($r^2$SCAN) functional is used for its improved accuracy and elimination of Hubbard U values that are necessary to describe transition metal oxide systems [38,39]. Calculation parameters were largely unchanged from defaults of the Materials Project *MPScanRelaxSet* for $r^2$SCAN calculations, enabling our calculations to be compatible with their extensively populated materials database for decomposition reaction energies [40,41]. A k-point mesh of 8x8x8 was used for the 2-atom rock salt unit cell and scaled linearly with the size of the supercell. Special quasi-random structures (SQSs) of 4x4x4 supercells (128-atoms) were generated using the Integrated Cluster Expansion Toolkit. [42,43] Due to supercell size constraints, the SQS for $Mg_{1/5}Co_{1/5}Ni_{1/5}Cu_{1/5}Zn_{1/5}O$ is slightly non-equimolar with 12 Zn atoms and 13 atoms for Mg, Co, Ni, and Cu; all four-component derivatives are equimolar. Antiferromagnetic ordering along the 111 direction is initialized for all rock salt calculations as observed previously for the parent HEO composition [23]. Competing phase calculations are performed with ferromagnetic ordering for consistency with the Materials Project. Calculations were managed and analyzed using Pymatgen and Sumo packages [44,45]. Optical calculations are performed using the independent particle approximation for calculating the frequency-dependent dielectric matrix in which the number of bands was doubled to ensure convergence of the energy spectrum.

# References:

[1] C.M. Rost, E. Sachet, T. Borman, A. Moballegh, E.C. Dickey, D. Hou, J.L. Jones, S. Curtarolo, J.-P. Maria, Entropy-stabilized oxides, Nature Communications 6 (2015) 8485. https://doi.org/10.1038/ncomms9485.

[2] G.N. Kotsonis, S.S.I. Almishal, F. Marques dos Santos Vieira, V.H. Crespi, I. Dabo, C.M. Rost, J.-P. Maria, High-entropy oxides: Harnessing crystalline disorder for emergent functionality, Journal of the American Ceramic Society 106 (2023) 5587–5611. https://doi.org/10.1111/jace.19252.

[3] G.N. Kotsonis, C.M. Rost, D.T. Harris, J.-P. Maria, Epitaxial entropy-stabilized oxides: growth of chemically diverse phases via kinetic bombardment, MRS Communications 8 (2018) 1371–1377. https://doi.org/10.1557/mrc.2018.184.

[4] G.N. Kotsonis, P.B. Meisenheimer, L. Miao, J. Roth, B. Wang, P. Shafer, R. Engel-Herbert, N. Alem, J.T. Heron, C.M. Rost, J.-P. Maria, Property and cation valence engineering in entropy-stabilized oxide thin films, Phys. Rev. Mater. 4 (2020) 100401. https://doi.org/10.1103/PhysRevMaterials.4.100401.

[5] S.S. Almishal, L. Miao, Y. Tan, G.N. Kotsonis, J.T. Sivak, N. Alem, L.-Q. Chen, V.H. Crespi, I. Dabo, C.M. Rost, Order evolution from a high-entropy matrix: understanding and predicting paths to low temperature equilibrium, arXiv Preprint arXiv:2404.15708 (2024).

[6] G.N. Kotsonis, S.S.I. Almishal, L. Miao, M.K. Caucci, G.R. Bejger, S.V.G. Ayyagari, T.W. Valentine, B.E. Yang, S.B. Sinnott, C.M. Rost, N. Alem, J.-P. Maria, Fluorite-structured high-entropy oxide sputtered thin films from bixbyite target, Applied Physics Letters 124 (2024) 171901. https://doi.org/10.1063/5.0201419.

[7] A.R. Mazza, J. Yan, S. Middey, J.S. Gardner, A.-H. Chen, M. Brahlek, T.Z. Ward, Embracing Disorder in Quantum Materials Design, (2024). https://doi.org/10.48550/arXiv.2402.18379.

[8] P.B. Meisenheimer, T.J. Kratofil, J.T. Heron, Giant Enhancement of Exchange Coupling in Entropy-Stabilized Oxide Heterostructures, Scientific Reports 7 (2017) 13344. https://doi.org/10.1038/s41598-017-13810-5.

[9] P.B. Meisenheimer, L.D. Williams, S.H. Sung, J. Gim, P. Shafer, G.N. Kotsonis, J.-P. Maria, M. Trassin, R. Hovden, E. Kioupakis, J.T. Heron, Magnetic frustration control through tunable stereochemically driven disorder in entropy-stabilized oxides, Phys. Rev. Mater. 3 (2019) 104420. https://doi.org/10.1103/PhysRevMaterials.3.104420.

[10] R. Witte, A. Sarkar, R. Kruk, B. Eggert, R.A. Brand, H. Wende, H. Hahn, High-entropy oxides: An emerging prospect for magnetic rare-earth transition metal perovskites, Phys. Rev. Mater. 3 (2019) 034406. https://doi.org/10.1103/PhysRevMaterials.3.034406.

[11] M.P. Jimenez-Segura, T. Takayama, D. Bérardan, A. Hoser, M. Reehuis, H. Takagi, N. Dragoe, Long-range magnetic ordering in rocksalt-type high-entropy oxides, Applied Physics Letters 114 (2019) 122401. https://doi.org/10.1063/1.5091787.

[12] X. Wang, H. Singh, M. Nath, K. Lagemann, K. Page, Excellent Bifunctional Oxygen Evolution and Reduction Electrocatalysts (5A1/5)Co2O4 and Their Tunability, ACS Mater. Au (2024). https://doi.org/10.1021/acsmaterialsau.3c00088.

[13] A. Sarkar, C. Loho, L. Velasco, T. Thomas, S.S. Bhattacharya, H. Hahn, R. Djenadic, Multicomponent equiatomic rare earth oxides with a narrow band gap and associated praseodymium multivalency, Dalton Trans. 46 (2017) 12167–12176. https://doi.org/10.1039/C7DT02077E.

[14] P. Edalati, Q. Wang, H. Razavi-Khosroshahi, M. Fuji, T. Ishihara, K. Edalati, Photocatalytic hydrogen evolution on a high-entropy oxide, Journal of Materials Chemistry A 8 (2020) 3814–3821.

[15] Zs. Rak, C.M. Rost, M. Lim, P. Sarker, C. Toher, S. Curtarolo, J.-P. Maria, D.W. Brenner, Charge compensation and electrostatic transferability in three entropy-stabilized oxides: Results from density functional theory calculations, Journal of Applied Physics 120 (2016) 095105. https://doi.org/10.1063/1.4962135.

[16] M. Anandkumar, P.M. Bagul, A.S. Deshpande, Structural and luminescent properties of Eu3+ doped multi-principal component Ce0. 2Gd0. 2Hf0. 2La0. 2Zr0. 2O2 nanoparticles, Journal of Alloys and Compounds 838 (2020) 155595.

[17] A.D. Dupuy, X. Wang, J.M. Schoenung, Entropic phase transformation in nanocrystalline high entropy oxides, Materials Research Letters 7 (2019) 60–67. https://doi.org/10.1080/21663831.2018.1554605.

[18] V. Jacobson, K. Gann, M. Sanders, G. Brennecka, Densification of the entropy stabilized oxide (Mg0.2Co0.2Ni0.2Cu0.2Zn0.2)O, Journal of the European Ceramic Society 42 (2022) 4328–4334. https://doi.org/10.1016/j.jeurceramsoc.2022.04.017.

[19] C.M. Rost, D.L. Schmuckler, C. Bumgardner, M.S. Bin Hoque, D.R. Diercks, J.T. Gaskins, J.-P. Maria, G.L. Brennecka, X. Li, P.E. Hopkins, On the thermal and mechanical properties of Mg0.2Co0.2Ni0.2Cu0.2Zn0.2O across the high-entropy to entropy-stabilized transition, APL Materials 10 (2022) 121108. https://doi.org/10.1063/5.0122775.

[20] M. Webb, M. Gerhart, S. Baksa, S. Gelin, A.-R. Ansbro, P.B. Meisenheimer, T. Chiang, J.-P. Maria, I. Dabo, C.M. Rost, J.T. Heron, High temperature stability of entropy-stabilized oxide (MgCoNiCuZn)0.2O in air, Applied Physics Letters 124 (2024) 151904. https://doi.org/10.1063/5.0199076.

[21] Zs. Rák, J.-P. Maria, D.W. Brenner, Evidence for Jahn-Teller compression in the (Mg, Co, Ni, Cu, Zn)O entropy-stabilized oxide: A DFT study, Materials Letters 217 (2018) 300–303. https://doi.org/10.1016/j.matlet.2018.01.111.

[22] D. Berardan, A.K. Meena, S. Franger, C. Herrero, N. Dragoe, Controlled Jahn-Teller distortion in (MgCoNiCuZn)O-based high entropy oxides, Journal of Alloys and Compounds 704 (2017) 693–700. https://doi.org/10.1016/j.jallcom.2017.02.070.

[23] N.J. Usharani, A. Bhandarkar, S. Subramanian, S.S. Bhattacharya, Antiferromagnetism in a nanocrystalline high entropy oxide (Co,Cu,Mg,Ni,Zn)O: Magnetic constituents and surface anisotropy leading to lattice distortion, Acta Materialia 200 (2020) 526–536. https://doi.org/10.1016/j.actamat.2020.09.034.

[24] C.J. Bartel, A.W. Weimer, S. Lany, C.B. Musgrave, A.M. Holder, The role of decomposition reactions in assessing first-principles predictions of solid stability, Npj Comput Mater 5 (2019) 4. https://doi.org/10.1038/s41524-018-0143-2.

[25] C.J. Bartel, Review of computational approaches to predict the thermodynamic stability of inorganic solids, J Mater Sci 57 (2022) 10475–10498. https://doi.org/10.1007/s10853-022-06915-4.

[26] D. Bérardan, S. Franger, D. Dragoe, A.K. Meena, N. Dragoe, Colossal dielectric constant in high entropy oxides, Physica Status Solidi (RRL)–Rapid Research Letters 10 (2016) 328–333.

[27] S. Swathilakshmi, R. Devi, G. Sai Gautam, Performance of the r2SCAN Functional in Transition Metal Oxides, J. Chem. Theory Comput. 19 (2023) 4202–4215. https://doi.org/10.1021/acs.jctc.3c00030.

[28] M. Moździerz, J. Dąbrowa, A. Stępień, M. Zajusz, M. Stygar, W. Zając, M. Danielewski, K. Świerczek, Mixed ionic-electronic transport in the high-entropy (Co,Cu,Mg,Ni,Zn)1-xLixO oxides, Acta Materialia 208 (2021) 116735. https://doi.org/10.1016/j.actamat.2021.116735.

[29] R.J. Spurling, S.S.I. Almishal, J. Casamento, J. Hayden, R. Spangler, M. Marakovits, A. Hossain, M. Lanagan, J.-P. Maria, Dielectric Properties of Disordered A6B2O17 (A = Zr; B = Nb, Ta) Phases, (2024). https://doi.org/10.48550/arXiv.2405.03527.

[30] F. Giustino, Materials modelling using density functional theory: properties and predictions, Oxford University Press, 2014.

[31] H.J. Kulik, Perspective: Treating electron over-delocalization with the DFT+U method, The Journal of Chemical Physics 142 (2015) 240901. https://doi.org/10.1063/1.4922693.

[32] Y. Pu, D. Moseley, Z. He, K.C. Pitike, M.E. Manley, J. Yan, V.R. Cooper, V. Mitchell, V.K. Peterson, B. Johannessen, R.P. Hermann, P. Cao, (Mg,Mn,Fe,Co,Ni)O: A rocksalt high-entropy oxide containing divalent Mn and Fe, Sci. Adv. 9 (2023) eadi8809. https://doi.org/10.1126/sciadv.adi8809.

[33] S. Divilov, H. Eckert, D. Hicks, C. Oses, C. Toher, R. Friedrich, M. Esters, M.J. Mehl, A.C. Zettel, Y. Lederer, E. Zurek, J.-P. Maria, D.W. Brenner, X. Campilongo, S. Filipović, W.G. Fahrenholtz, C.J. Ryan, C.M. DeSalle, R.J. Crealese, D.E. Wolfe, A. Calzolari, S. Curtarolo, Disordered enthalpy–entropy descriptor for high-entropy ceramics discovery, Nature 625 (2024) 66–73. https://doi.org/10.1038/s41586-023-06786-y.

[34] F. Urbach, The Long-Wavelength Edge of Photographic Sensitivity and of the Electronic Absorption of Solids, Phys. Rev. 92 (1953) 1324–1324. https://doi.org/10.1103/PhysRev.92.1324.

[35] B. Sundman, U.R. Kattner, C. Sigli, M. Stratmann, R. Le Tellier, M. Palumbo, S.G. Fries, The OpenCalphad thermodynamic software interface, Computational Materials Science 125 (2016) 188–196.

[36] R. Otis, Z.-K. Liu, pycalphad: CALPHAD-based Computational Thermodynamics in Python, Journal of Open Research Software 5 (2017) 1–1.

[37] M. Hillert, The compound energy formalism, Journal of Alloys and Compounds 320 (2001) 161–176.

[38] J.W. Furness, A.D. Kaplan, J. Ning, J.P. Perdew, J. Sun, Accurate and Numerically Efficient $r^2$SCAN Meta-Generalized Gradient Approximation, J. Phys. Chem. Lett. 11 (2020) 8208–8215. https://doi.org/10.1021/acs.jpclett.0c02405.

[39] M. Kothakonda, A.D. Kaplan, E.B. Isaacs, C.J. Bartel, J.W. Furness, J. Ning, C. Wolverton, J.P. Perdew, J. Sun, Testing the $r^2$SCAN density functional for the thermodynamic stability of solids with and without a van der Waals correction, (2022). http://arxiv.org/abs/2208.02841 (accessed September 20, 2022).

[40] A. Jain, S.P. Ong, G. Hautier, W. Chen, W.D. Richards, S. Dacek, S. Cholia, D. Gunter, D. Skinner, G. Ceder, Commentary: The Materials Project: A materials genome approach to accelerating materials innovation, APL Materials 1 (2013).

[41] R. Kingsbury, A.S. Gupta, C.J. Bartel, J.M. Munro, S. Dwaraknath, M. Horton, K.A. Persson, Performance comparison of $r^2$SCAN and SCAN metaGGA density functionals for solid materials via an automated, high-throughput computational workflow, Phys. Rev. Materials 6 (2022) 013801. https://doi.org/10.1103/PhysRevMaterials.6.013801.

[42] M. Ångqvist, W.A. Muñoz, J.M. Rahm, E. Fransson, C. Durniak, P. Rozyczko, T.H. Rod, P. Erhart, ICET – A Python Library for Constructing and Sampling Alloy Cluster Expansions, Advanced Theory and Simulations 2 (2019) 1900015. https://doi.org/10.1002/adts.201900015.

[43] A. Zunger, S.-H. Wei, L. Ferreira, J.E. Bernard, Special quasirandom structures, Physical Review Letters 65 (1990) 353.

[44] A.M. Ganose, A.J. Jackson, D.O. Scanlon, sumo: Command-line tools for plotting and analysis of periodic* ab initio* calculations, Journal of Open Source Software 3 (2018) 717.

[45] S.P. Ong, W.D. Richards, A. Jain, G. Hautier, M. Kocher, S. Cholia, D. Gunter, V.L. Chevrier, K.A. Persson, G. Ceder, Python Materials Genomics (pymatgen): A robust, open-source python library for materials analysis, Computational Materials Science 68 (2013) 314–319.

# Supporting Information

**Table S1.** Prescribed thermodynamic data for pseudo-binary phase diagram calculations.

| Phase and Sublattice Model | Species with Prescribed Chemical Potential | Ref. |
|---|---|---|
| L (liquid) $(Co^{2+},Cu^{+},Mg^{2+},Ni^{2+},Zn^{2+})_1(O^{2-})_1$ | CoO | [1] |
| | $Cu_2O$ | [2] |
| | MgO | [3] |
| | NiO | [3] |
| | ZnO | [4] |
| J14 rock salt (halite) $(Co^{2+},Cu^{2+},Mg^{2+},Ni^{2+},Zn^{2+})_1(O^{2-})_1$ | CoO | [1] |
| | CuO | [5] |
| | MgO | [3] |
| | NiO | [3] |
| | ZnO | [4] |
| CuO (tenorite) $(Cu^{2+},Co^{2+})_1(O^{2-})_1$ | CuO | [5] |
| | CoO | |
| $Cu_2O$ (cuprite) | $Cu_2O$ | [5] |
| $Co_3O_4$ (spinel) $(Co^{2+},Co^{3+})_1(Co^{2+},Co^{3+})_2(O^{2-})_4$ | $(Co^{2+})_1(Co^{3+})_2(O^{2-})_4$ (normal spinel) | [1] |
| | $(Co^{3+})_1(Co^{2+},Co^{3+})_2(O^{2-})_4$ (inverse spinel) | |
| ZnO (wurtzite) $(Zn^{2+},Mg^{2+},Co^{2+},Ni^{2+})_1(O^{2-})_1$ | ZnO | [4] |
| | MgO | [4] |
| | CoO | [6] |
| | NiO | [7] |
| $MgCu_2O_3$ (Güggenite) | $MgCu_2O_3$ | [8] |
| $Co_2CuO_3$ | $Co_2CuO_3$ | [5] |
| $O_2$ (gas) | $O_2$ | [1] |

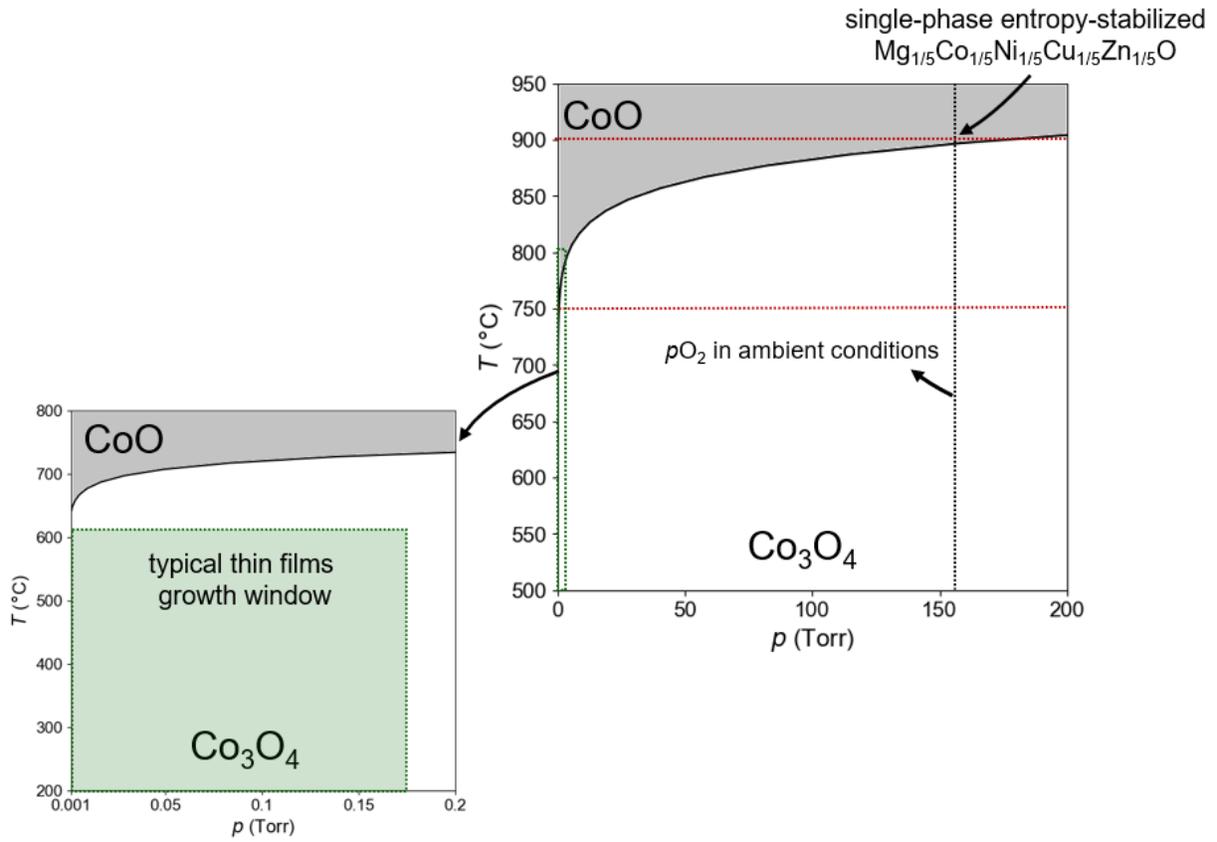

**Figure S1.** Co-O system showing CoO-Co$_3$O$_4$ stability with temperature and oxygen partial pressure.

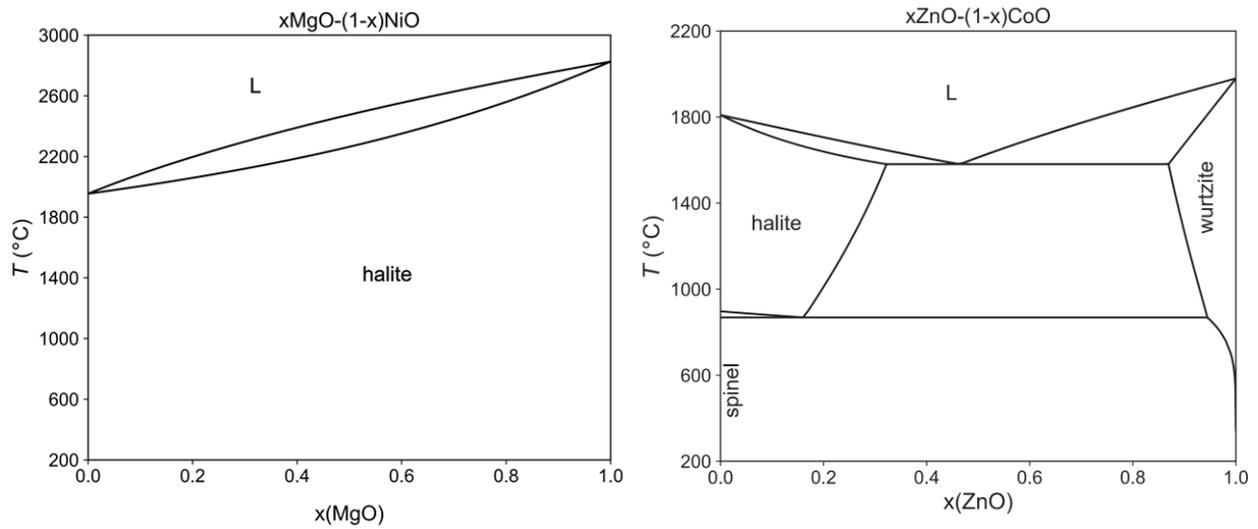

**Figure S2.** Binary phase diagrams for MgO and NiO with complete solubility and CoO and ZnO with no apparent solubility.

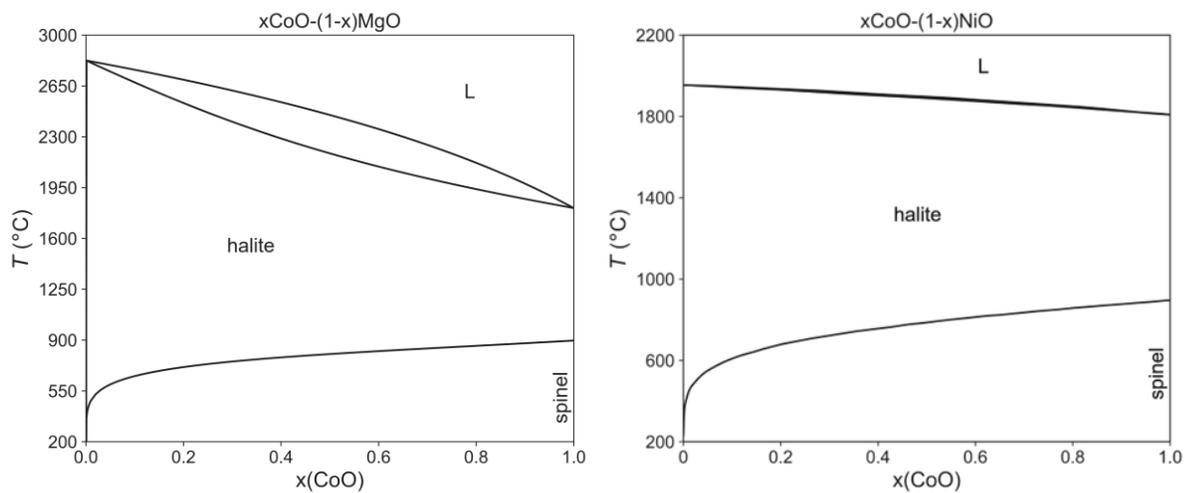

**Figure S3.** Binary phase diagrams for Co solubility in MgO and NiO.

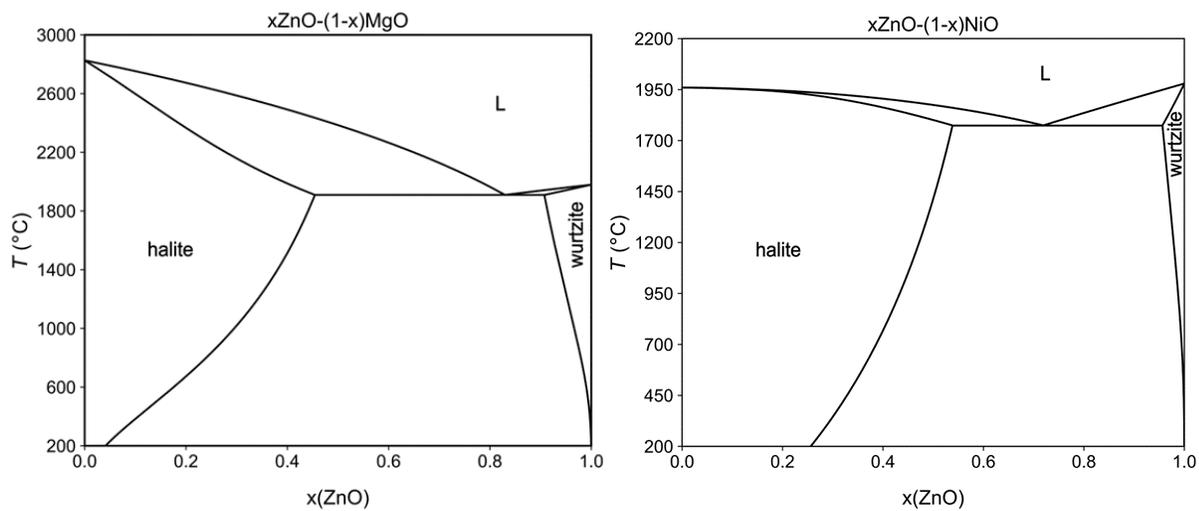

**Figure S4.** Binary phase diagrams for ZnO solubility in MgO and NiO.

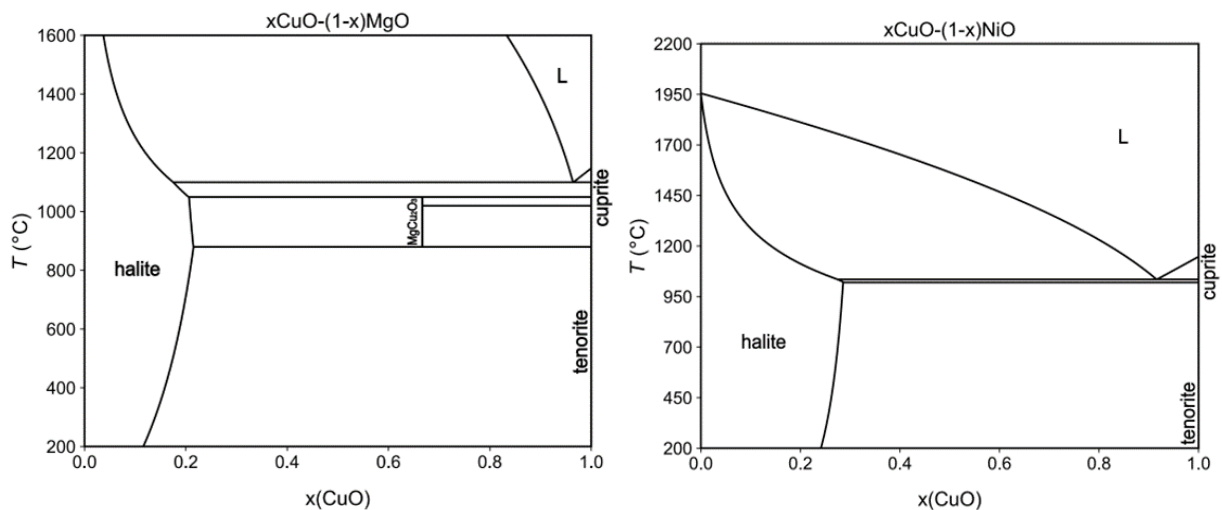

**Figure S5**. Binary phase diagrams for CuO solubility in MgO and NiO.

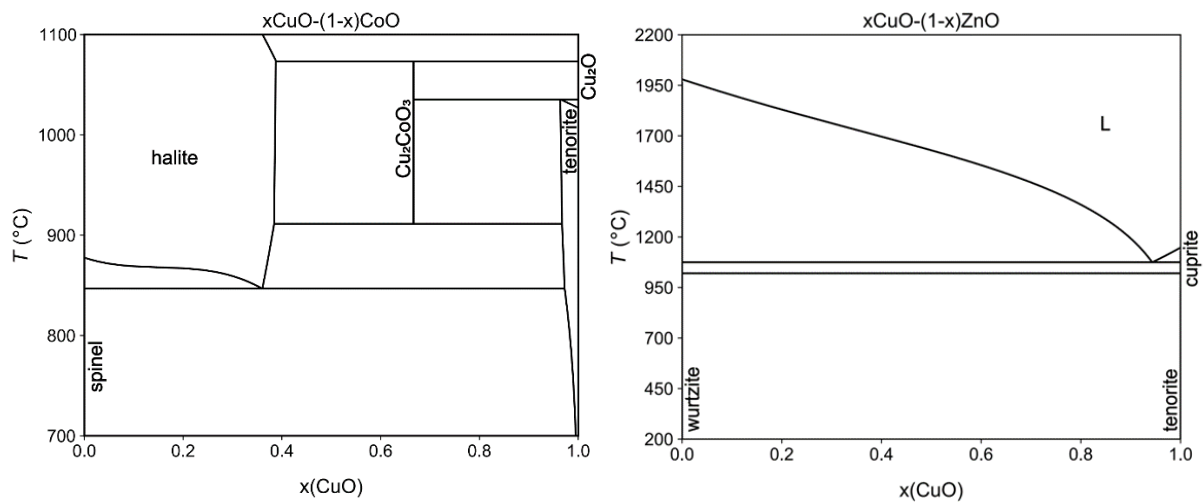

**Figure S6.** Binary phase diagrams for CuO with $Co_xO_y$ and ZnO.

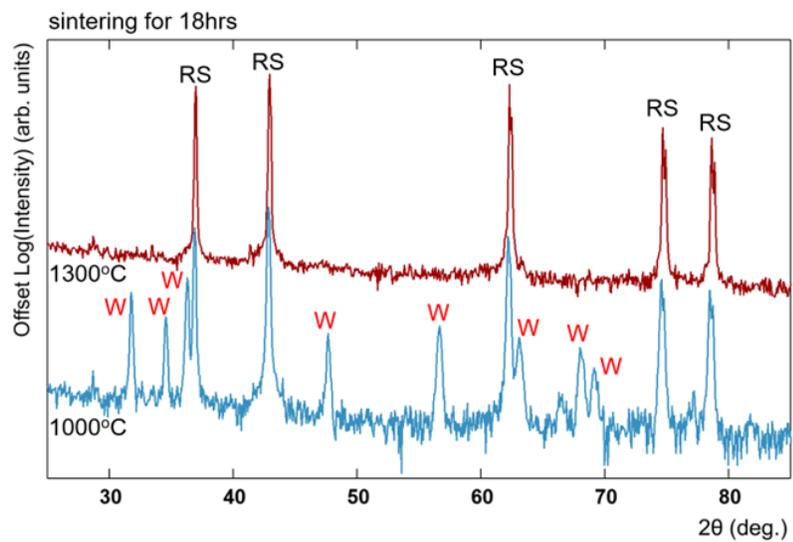

**Figure S7.** X-ray diffraction patterns of two bulk ceramics mixed from MgO, NiO, and ZnO to form stoichiometric $Mg_{1/3}Ni_{1/3}Zn_{1/3}O$ and are sintered at 1000 °C (bottom) and at 1300 °C (top) for 18 hours. Intermediate temperatures are not shown for clarity. The wurtzite phase signatures start disappearing around 1300 °C.

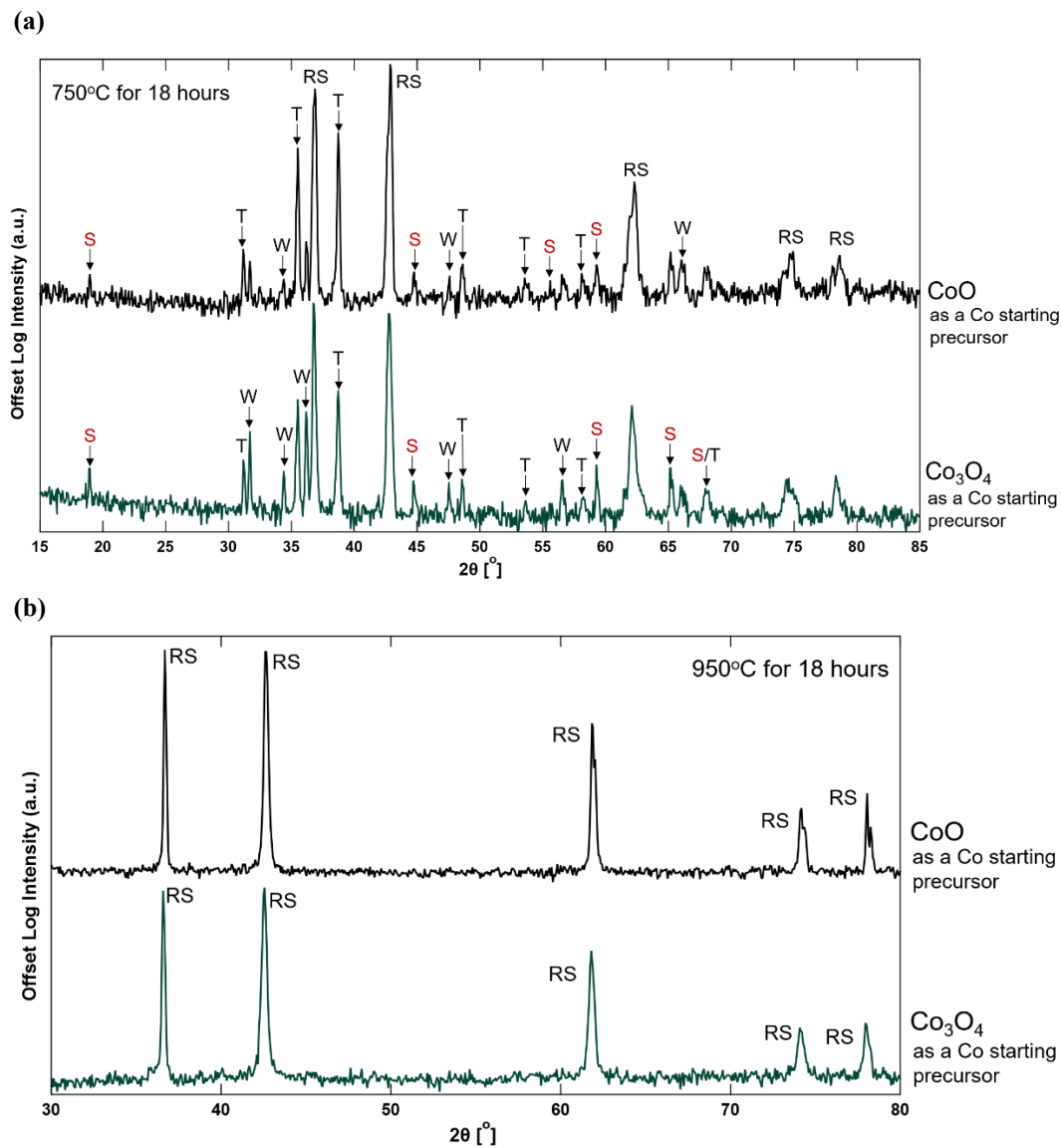

**Figure S8.** (a) X-ray diffraction patterns of two bulk ceramics mixed from CoO powders (top) and $Co_3O_4$ powders (bottom). Each powder set result in stoichiometric $Mg_{1/5}Co_{1/5}Ni_{1/5}Cu_{1/5}Zn_{1/5}O$ and are sintered at 750 °C for 18 hours. (b) X-ray diffraction patterns of two bulk ceramics mixed from CoO powders (top) and $Co_3O_4$ powders (bottom). Each powder set result in single-phase rock salt stoichiometric $Mg_{1/5}Co_{1/5}Ni_{1/5}Cu_{1/5}Zn_{1/5}O$ and are sintered at 950 °C for 18 hours.

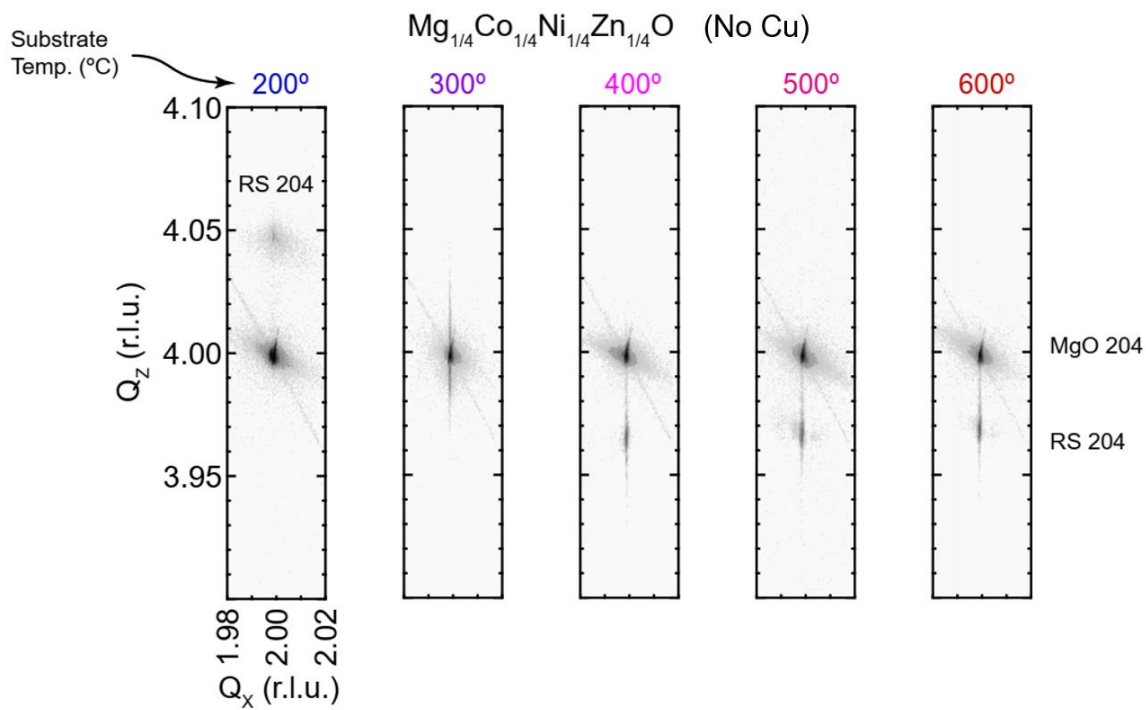

**Figure S9.** Reciprocal space maps of $Mg_{1/4}Co_{1/4}Ni_{1/4}Zn_{1/4}O$ (No Cu) and MgO 204 reflections, indicating commensurate film growth. Substrate temperature during film growth is indicated near each plot.

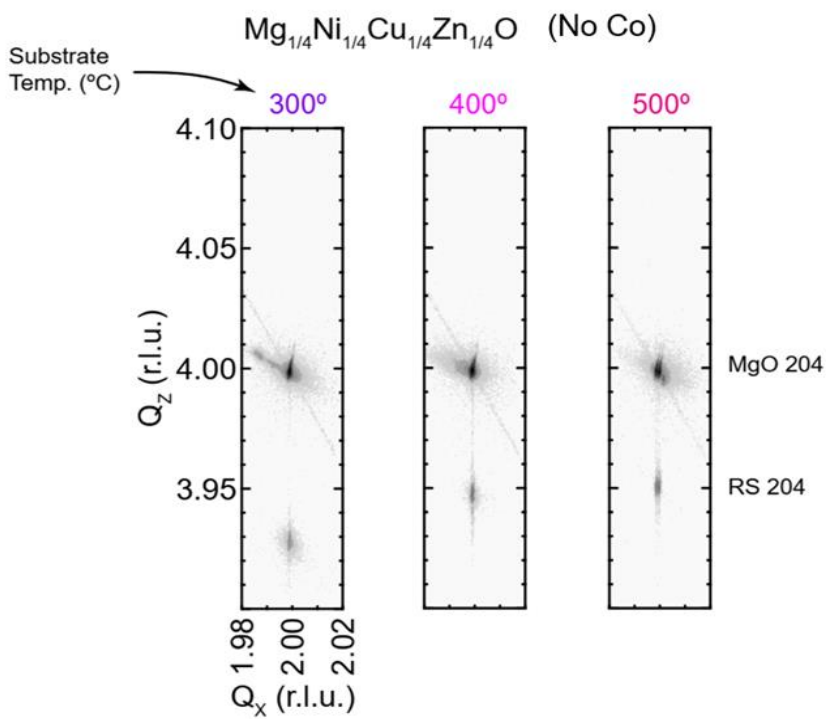

**Figure S10.** Reciprocal space maps of $Mg_{1/4}Ni1/4Zn_{1/4}Cu_{1/4}O$ (No Co) and MgO 204 reflections, indicating commensurate film growth. Substrate temperature during film growth is indicated near each plot.

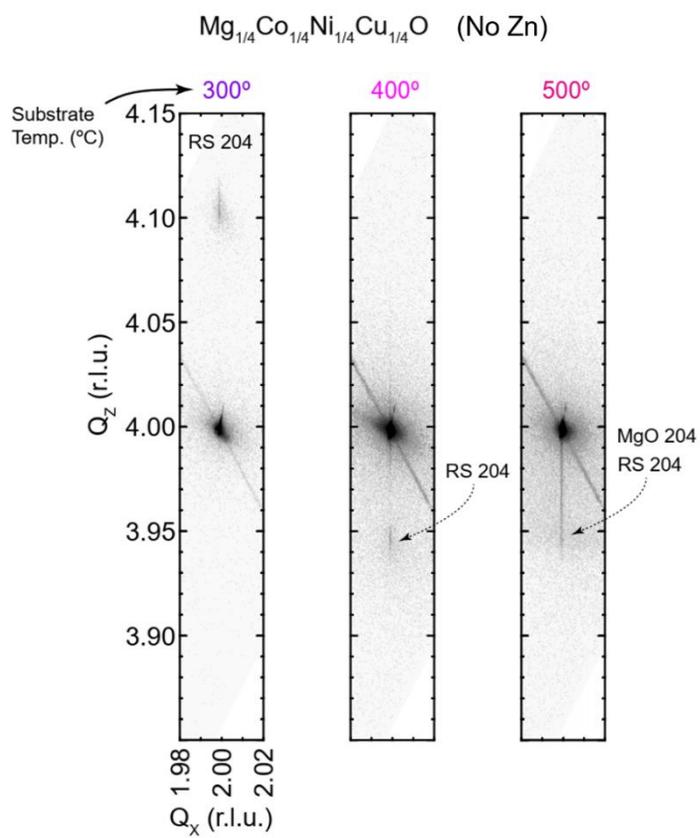

**Figure S11.** Reciprocal space maps of $Mg_{1/4}Co_{1/4}Ni_{1/4}Cu_{1/4}O$ (No Zn) and MgO 204 reflections, indicating commensurate film growth. Substrate temperature during film growth is indicated near each plot.

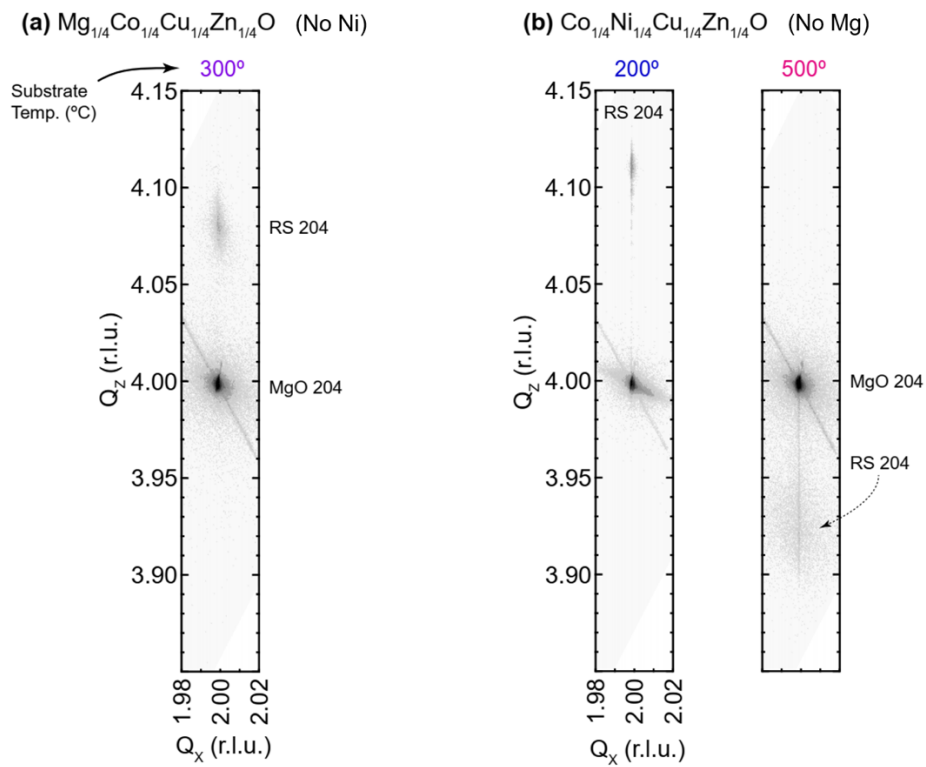

**Figure S12.** Reciprocal space maps of Mg$_{1/4}$Co$_{1/4}$Ni$_{1/4}$Cu$_{1/4}$O (No Zn) and MgO 204 reflections, indicating commensurate film growth. Figure S11: Reciprocal space maps of: (a) Mg$_{1/4}$Co$_{1/4}$Cu$_{1/4}$Zn$_{1/4}$O (J14-Ni); (b) Co$_{1/4}$Ni$_{1/4}$Cu$_{1/4}$Zn$_{1/4}$O (J14-Mg); and MgO 204 reflections. Film 204 Q$_X$ coordinates indicate commensurate film growth for both compositions. Substrate temperature during film growth is indicated near each plot.

**Table S2.** Decomposition reactions corresponding to decomposition enthalpies for parent $Mg_{1/5}Co_{1/5}Ni_{1/5}Cu_{1/5}Zn_{1/5}O$ and its four-component derivatives. All competing phase energies are extracted from the Materials Project database of r²SCAN calculations.

| Composition | Decomposition Reaction |
| --- | --- |
| $Mg_{1/5}Co_{1/5}Ni_{1/5}Cu_{1/5}Zn_{1/5}O$ | 0.41 ($MgNiO_2$) + 0.36 ($ZnCo_2O_4$) + 0.10 ($CuO$) + 0.09 ($ZnO$) + 0.05 ($Cu$) |
| $Co_{1/4}Ni_{1/4}Cu_{1/4}Zn_{1/4}O$ | 0.44 ($CoNi_2O_4$) + 0.22 ($ZnCo_2O_4$) + 0.06 ($CuO$) + 0.19 ($ZnO$) + 0.09 ($Cu$) |
| $Mg_{1/4}Ni_{1/4}Cu_{1/4}Zn_{1/4}O$ | 0.50 ($MgNiO_2$) + 0.25 ($CuO$) + 0.25 ($ZnO$) |
| $Mg_{1/4}Co_{1/4}Cu_{1/4}Zn_{1/4}O$ | 0.44 ($ZnCo_2O_4$) + 0.12 ($ZnO$) + 0.25 ($MgO$) + 0.06 ($Cu$) + 0.12 ($CuO$) |
| $Mg_{1/4}Co_{1/4}Ni_{1/4}Zn_{1/4}O$ | 0.25 ($MgNiO_2$) + 0.44 ($ZnCo_2O_4$) + 0.12 ($ZnO$) + 0.12 ($MgO$) + 0.06 ($Ni$) |
| $Mg_{1/4}Co_{1/4}Ni_{1/4}Cu_{1/4}O$ | 0.50 ($MgNiO_2$) + 0.17 ($CuO$) + 0.29 ($Co_3O_4$) + 0.04 ($Cu$) |

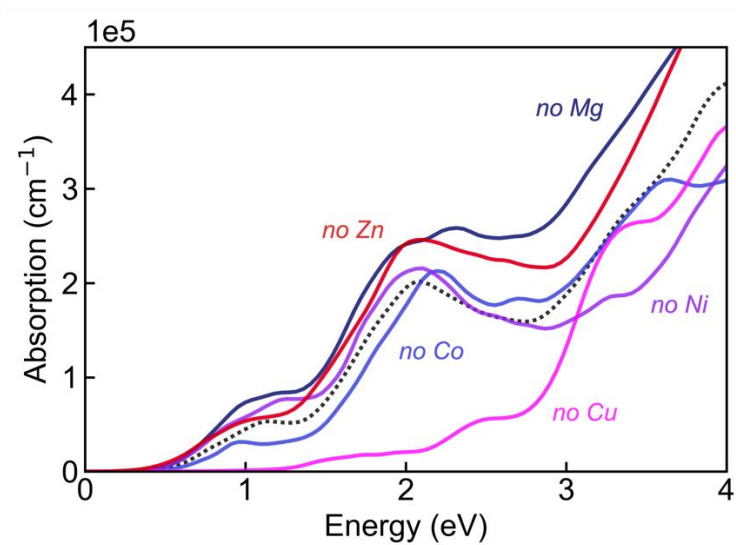

**Figure S13.** Optical absorption of parent $Mg_{1/5}Co_{1/5}Ni_{1/5}Cu_{1/5}Zn_{1/5}O$ and its four-component derivatives. $Mg_{1/5}Co_{1/5}Ni_{1/5}Cu_{1/5}Zn_{1/5}O$ is shown as a black dotted line.

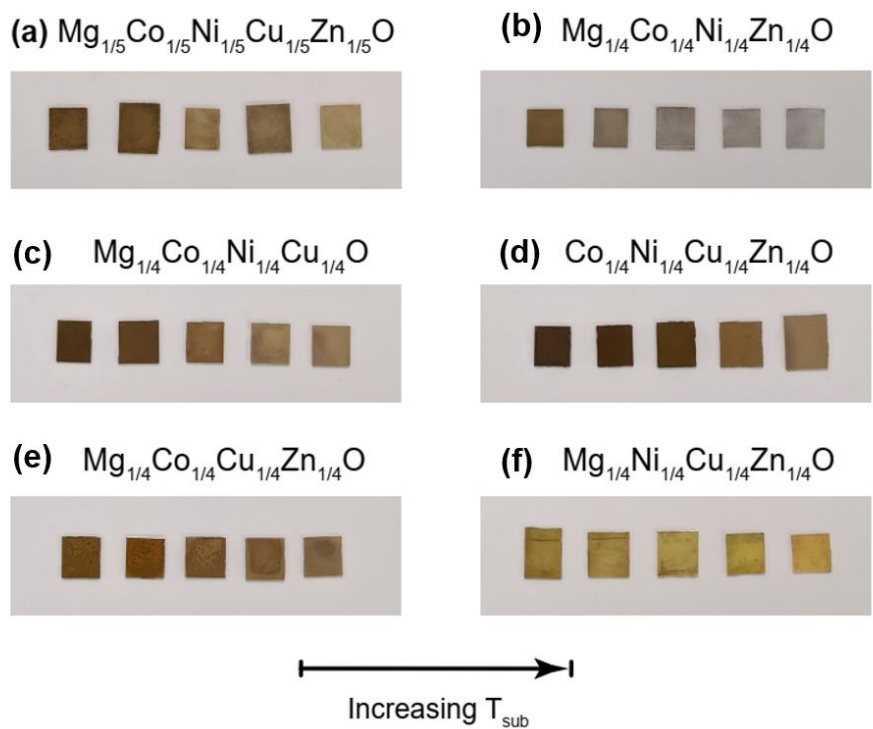

**Figure S14.** Photographs of films grown on [001]-MgO at 200–500 °C: (a) $Mg_{1/5}Co_{1/5}Ni_{1/5}Cu_{1/5}O$; (b) $Mg_{1/4}Co_{1/4}Ni_{1/4}Zn_{1/4}O$ (no Cu); (c) $Mg_{1/4}Co_{1/4}Ni_{1/4}Cu_{1/4}O$ (no Zn); (d) $Co_{1/4}Ni_{1/4}Cu_{1/4}Zn_{1/4}O$ (no Mg); (e) $Mg_{1/4}Co_{1/4}Cu_{1/4}Zn_{1/4}O$ (no Ni); (f) $Mg_{1/4}Ni_{1/4}Cu_{1/4}Zn_{1/4}O$ (No Co). Note that the change in absorption coefficient ($\alpha = 4\pi k \lambda$) and band gap indicated by ellipsometry models is also evident by visual inspection.

## References:

[1] M. Chen, B. Hallstedt, L.J. Gauckler, Thermodynamic assessment of the Co-O system, Journal of Phase Equilibria 24 (2003) 212–227. https://doi.org/10.1361/105497103770330514.

[2] B. Hallstedt, D. Risold, L.J. Gauckler, Thermodynamic Assessment of the Copper-Oxygen System, Joumal of Phase Equilibria 15 (1994) 483–499. https://doi.org/10.1007/BF02649399.

[3] V. Prostakova, J. Chen, E. Jak, S.A. Decterov, Experimental study and thermodynamic optimization of the CaO-NiO, MgO-NiO and NiO-SiO2 systems, Calphad: Computer Coupling of Phase Diagrams and Thermochemistry 37 (2012) 1–10. https://doi.org/10.1016/j.calphad.2011.12.009.

[4] J. Jeon, D. Lindberg, Thermodynamic optimization and phase equilibria study of the MgO–ZnO, CaO–ZnO, and CaO–MgO–ZnO systems, Ceramics International 49 (2023) 12736–12744. https://doi.org/10.1016/j.ceramint.2022.12.138.

[5] L.A. Zabdyr, O.B. Fabrichnaya, Phase equilibria in the cobalt oxide-copper oxide system, Journal of Phase Equilibria 23 (2002) 149–155. https://doi.org/10.1361/1054971023604161.

[6] C. Ma, A. Navrotsky, Thermodynamics of the CoO-ZnO system at bulk and nanoscale, Chemistry of Materials 24 (2012) 2311–2315. https://doi.org/10.1021/cm3005198.

[7] P.K. Davies, A. Navrotsky, Thermodynamics of solid solution formation in NiO-MgO and NiO-ZnO, Journal of Solid State Chemistry 38 (1981) 264–276. https://doi.org/10.1016/0022-4596(81)90044-X.

[8] J. Assal, B. Hallstedt, L.J. Gauckler, Thermodynamic Evaluation of the Mg-Cu-O System, Zeitschrift Für Metallkunde 87 (1996) 568–573. https://doi.org/10.1515/ijmr-1996-870709.